\documentclass[usenatbib]{mn2e}
\usepackage{graphicx}
\usepackage{url}
\usepackage{amsmath}
\usepackage{threeparttable}
\usepackage{url}
\usepackage{hyperref}
\usepackage{caption3}
\usepackage[all]{hypcap}        
\usepackage[capitalise]{cleveref}
\crefformat{equation}{Eq.~#2#1#3}

\usepackage{xcolor} 
\definecolor{darkblue}{rgb}{0,0,0.5}
\hypersetup{
  colorlinks   = true, 
  urlcolor     = blue, 
  linkcolor    = blue, 
  citecolor    = darkblue   
}

\newcommand{\msun}{\mbox{\,M$_\odot$}}        
\newcommand{\lbol}{$\mathrm{L}_\mathrm{bol}$}                       
\newcommand{\Mbol}{$\mathrm{M}_\mathrm{bol}$}                       
\newcommand{\teff}{$\mathrm{T}_\mathrm{eff}$} 
\newcommand{\kms}{km~s$^{-1}$}
\newcommand{\ca}{Ca-rich transient}
\newcommand{\cas}{Ca-rich transients}

\newcommand{\simlt}{{\small\raisebox{-0.6ex}{$\,\stackrel{\raisebox{-.2ex}{$\textstyle <$}}{\sim}\,$}}}
\newcommand{\simgt}{{\small\raisebox{-0.6ex}{$\,\stackrel{\raisebox{-.2ex}{$\textstyle >$}}{\sim}\,$}}}

\begin{document}

\title[Calcium-rich transients]{
The progenitors of calcium-rich transients are not formed in situ
\thanks{Based on observations made with ESO Telescopes at the Paranal Observatory under programme ID 092.D-0420}}

\author[Lyman et al.]
{\parbox{\textwidth}{J. D. Lyman$^1$\thanks{E-mail: J.D.Lyman@warwick.ac.uk}, 
A. J. Levan$^1$,
R. P. Church$^2$,
M. B. Davies$^2$,
N. R. Tanvir$^3$
}
\vspace{0.4cm}\\
$^1$Department of Physics, University of Warwick, Coventry CV4 7AL, UK\\
$^2$Lund Observatory, Department of Astronomy and Theoretical Physics, Box 43, SE-221 00 Lund, Sweden\\
$^3$Department of Physics and Astronomy, University of Leicester, Leicester, LE1 7RH, UK\\
}

\date{Accepted . Received ; in original form }

\pagerange{\pageref{firstpage}--\pageref{lastpage}} \pubyear{2014}

\maketitle

\label{firstpage}

\begin{abstract}
We present deep VLT and {\em HST} observations of the nearest examples of Ca-rich ``gap" transients -- rapidly evolving transient events, with a luminosity intermediate between novae and supernovae. These sources are frequently found at  large galactocentric offsets, and their progenitors remain mysterious. Our observations find no convincing underlying quiescent sources co-incident with the locations of these transients, allowing us to rule out a number of potential progenitor systems.  The presence of surviving massive-star binary companions (or other cluster members) are ruled out, providing an independent rejection of a massive star origin for these events. Dwarf satellite galaxies are disfavoured unless one invokes as yet unknown conditions that would be extremely favourable for their production in the lowest mass systems. Our limits also probe the majority of the globular cluster luminosity function, ruling out the presence of an underlying globular cluster population at high significance, and thus the possibility that they are created via dynamical interactions in dense globular cluster cores. 
Given the lack of underlying systems, previous progenitor suggestions have difficulty reproducing the remote locations of these transients, even when considering solely halo-borne progenitors.
Our preferred scenario is that \cas{} are high velocity, kicked systems, exploding at large distances from their natal site. Coupled with a long-lived progenitor system post-kick, this naturally explains the lack of association these transients have with their host stellar light, and the extreme host-offsets exhibited.
Neutron star -- white dwarf mergers may be a promising progenitor system in this scenario.
\end{abstract}


\section{Introduction}
\label{sect:intro}

In recent years, dedicated optical transient surveys have been monitoring larger and larger fractions of the night sky with ever faster cadence. Such a strategy is not only overwhelming previous detection numbers of well-established transients such as novae and supernovae (SNe), but is also revealing and expanding samples of fast and faint transients that do not appear to fit easily into existing classification schemes. One such example are `calcium-rich' transients \citep{filippenko03}, also termed `SN~2005E-like' events. As defined by \citet{kasliwal12}, \cas{} are fainter and evolve faster (photometrically and spectroscopically) than most SNe, whilst maintaining photospheric velocities typical of normal SNe. At peak \cas{} have luminosities intermediate between classical novae and SNe, and populate a `luminosity gap', where transients have typically been difficult to locate.
Their nebular spectra give the class its name since they are dominated by calcium, with a high Ca/O ratio (or lack of detected oxygen) compared to that of other SNe, for example \citet{perets10} found almost half of the ejecta of SN~2005E to be calcium. 
Besides the noticeable difference in evolution speed and peak luminosity, \cas{} can look similar to Type Ib SNe (SNe~Ib) during the early, photospheric evolution -- i.e. there is an absence of hydrogen in their spectra but they show strong helium features \citep{filippenko03,perets10}, although \citet{kasliwal12} leave a description of the photospheric spectra out of their defining characteristics since not all members have spectral coverage or even show clear evidence for helium \citep{sullivan11,valenti14}. 

The `luminosity gap' between classical novae and SNe may not exist much longer as it is being gradually populated by various peculiar transient events -- here 

we briefly discuss the distinction between \cas{} and other notable classes.
The SN-like photospheric velocities ($\sim$10000~\kms{}) of these transients distinguishes them from another `fast and faint' class of transient -- SN~2002cx-like transients or `Iax' \citep{li03,foley13} -- alongside differences in their host environments \citep{lyman13}. A further class are `.Ia' \citep{bildsten07}, so-named for their luminosity being around a tenth of that of SNe~Ia, with SNe~2002bj \citep{poznanski10} and 2010X \citep{kasliwal10} being prominent examples. They are, nevertheless, much brighter than \cas{}, and also decay much more rapidly.

The apparent similarity of the \ca{} SN~2005cz to a SN~Ib (i.e. showing helium in its photospheric spectra) led to the claim of \citet{kawabata10} that SN~2005cz was a core-collapse SN (CCSN) originating from an $\sim8-12$~\msun{} progenitor. \citet{perets10} and \citet{perets11}, however, showed for SNe~2005E and 2005cz respectively that there were no signatures of underlying recent or ongoing star formation and the environments were indeed consistent with being very old stellar populations. \citet{lyman13} presented an analysis of the environments of all \ca{} events\footnote{We define the entire sample of \cas{} as those presented in \citet{lyman13}, however we discuss the implications of contamination in \cref{sect:sampledef}.} by looking at the statistical distributions of host galaxy morphologies and associations to regions of recent star formation. Both galaxy morphologies (6/12 \ca{} are associated with E/S0 galaxies cf. none out of hundreds of CCSNe, \citealt{barbon99}\footnote{\url{http://heasarc.gsfc.nasa.gov/W3Browse/all/asiagosn.html}}) and a complete lack of association to star formation led the authors to argue for very long-lived progenitors.\footnote{A follow up analysis, presented in \citet{lyman14b}, further rules out progenitors with the lifetimes of CCSN progenitors, and favours ages $>$100~Myr.} This is consistent with the findings of \citet{yuan13}, who infer ages of the progenitors of Gyrs and find them to likely be metal-poor given their large host offsets. The galactocentric distances of \ca{} appear extreme when compared to normal SNe and can be tens of kpc away from their putative hosts in plane-of-sky alone \citep{kasliwal12}; one third show a projected offset of $>$~20~kpc (\cref{tab:offsets}). A plausible explanation could be that they are formed in globular clusters (GCs), where dynamical interactions create exotic systems at higher rate, per unit mass, than the field. Their apparent preference for large galactocentric distances and the extremely high calcium yield of their ejecta has been suggested as a means to alleviate the discrepancies between the high calcium abundance seen in intracluster media and that which can be produced from models relying on SNe~Ia and CCSNe alone \citep{mulchaey14}. This is evidence that these transients, although diminutive by other SN standards, may have an important role to play in the chemical evolution of large scale structure. A very recent study by \citet{perets14} has also suggested the contribution of the nucleosynthetic products of \cas{} to the observed Galactic 511 keV flux is significant -- providing an alternative to dark matter annihilation scenarios.

Recent effort into these peculiar transients has seen a number of progenitor models being proposed. Massive stars \citep{kawabata10} appear disfavoured by current observational analyses of the transients' locations, although relatively little study of the potential of such stars to produce faint CCSN I (e.g. fall-back SNe, \citealt{fryer09}; \citealt{moriya10}) has been done. A variant of the `double degenerate' scenario producing `.Ia' explosions \citep[e.g.][]{bildsten07} was proposed by \citet{perets10} to explain \cas{}. This involves the explosion of an accreted helium-rich layer (from a helium white dwarf; WD) on the surface of a carbon-oxygen WD, with simulations able to roughly reproduce the spectral and photometric evolution of SN~2005E, although some differences in peak luminosity and light curve evolution exist \citep{waldman11,sim12}. Such a progenitor would have the requisite long-lived nature and a preference for low metallicity \citep{waldman11}.

This paper provides further constraints on the nature of \cas{} by analysing the locations of nearby members of the class down to very deep limits and comparing with expectations for underlying systems. Details of the observations and the transients analysed are presented in \cref{sect:obs}, results are presented in \cref{sect:results} and discussed in the context of potential progenitor systems in \cref{sect:discuss}.

\begin{table}
\begin{threeparttable}
 \caption{Host-galaxy offsets of \cas{}}
 \begin{tabular}{llc}
\hline
Transient  & Host &  Offset\tnote{a}\\
           &      &  (kpc)\\
\hline
SN~2000ds  &  NGC~2768      &                 3              \\
SN~2001co  &  NGC~5559      &               6              \\
SN~2003H   &  NGC~2207/IC~2163  &                   6--9\tnote{b}  \\
SN~2003dg  &  UGC~6934         &            2              \\
SN~2003dr  &  NGC~5714         &            3              \\
SN~2005E   &  NGC~1032          &           23             \\
SN~2005cz  &  NGC~4589          &           2              \\
SN~2007ke  &  NGC~1129/MCG+07-07-003    &                 8--17\tnote{b} \\
PTF~09dav  &  Anon                 &    43             \\
SN~2010et  &  multiple\tnote{c} &  35--65\tnote{b}\\
PTF~11bij  &  IC~3956              &       34             \\
SN~2012hn  &  NGC~2272              &       8              \\
\end{tabular}
\label{tab:offsets}
\begin{tablenotes}
 \item [a]{Offsets were found using the transient's coordinates and galaxy centre coordinates and distances from NED.}
 \item [b]{More than one nearby galaxy -- the range encompasses the offsets from the nearest and furthest potential hosts.}
 \item [c]{\raggedright{CGCG\mbox{~}170-010/CGCG\mbox{~}170-011/SDSS\mbox{~}J171650.20 +313234.4}}
\end{tablenotes}
\end{threeparttable}
\end{table}

\section{Sample}
\label{sect:obs}

\subsection{New imaging and data reduction}

Two proximate ($\sim$30--35~Mpc) examples of the \cas{} sample were selected for deep optical imaging in order to detect (or put stringent limits upon) any underlying sources, these were SNe 2005E \citep{perets10} and 2012hn \citep{valenti14}. Imaging was obtained with the FORS2 instrument on the Very Large Telescope (VLT), details of these observations can be found in \cref{tab:obs}. Data were reduced and calibrated using recipes from the FORS2 pipeline, accessed from within the {\sc gasgano} framework.

Additionally, data from the Hubble Space Telescope (HST) data archive were obtained for SN~2003H, another nearby \ca{} at a distance of $\sim$35~Mpc \citep{li11c}. Details of the observations are provided in \cref{tab:obs2}. Two epochs were obtained, the first prior to the explosion of SN~2003H with WFPC2 (GO 6483, Elmegreen), and the latter post explosion utilizing the ACS/HRC
(GO 10272, Filippenko). The data were reduced via {\tt astrodrizzle} to 
produce final co-added, distortion corrected images. Given the limited dithering for each observation we retained the native pixel scale (0.1 arcsec for WFPC2 and 0.025 arcsec for ACS/HRC). For comparison to the earlier WFPC2 images we also drizzled the HRC images onto a larger pixel grid of 0.1 arcsec per pixel.

\subsection{Archival limits}

Two Ca-rich transients that have been analysed previously with deep limits on their locations are included in the sample of this paper. SN~2000ds was investigated as part of a SN progenitor detection study by \citet{maund05}, with no detection made to the limits of their observations. The environment of SN~2005cz was studied by \citet{perets11} as a means to constrain the progenitor of that event, with the authors finding the location to be composed of an old stellar population. We include the limits and results presented from these studies where appropriate in our analysis of \cas{}. 

Additionally, limits have also been presented for three PTF-discovered events by \citet{kasliwal12}, PTF 09dav, 10iuv (= SN~2010et) and 11bij. Given the larger distances to these transients, the depth of the images are comparatively shallow in terms of absolute magnitude. However, we note that no detections have been made underlying these events and thus only limits exist. These limits are qualitatively addressed in the discussion as appropriate for completeness' sake.

\begin{table}
 \caption{VLT observations of \cas{}}
 \begin{tabular}{lcccc}
\hline
Transient  & Observation & Filter & Exposure time & Seeing   \\
           &  Date       &        & (s)           & (arcsec) \\
\hline
SN~2005E      & 22 Jan 2014 & $R$    & 2400          & 0.8      \\ 
SN~2012hn     & 03 Feb 2014 & $B$    & 3600          & 0.8      \\
           & 07 Dec 2013 & $R$    & 3000          & 0.7      \\ 
\hline
 \end{tabular}
\label{tab:obs}
\end{table}

\begin{table}
 \caption{HST observations of SN~2003H}
 \begin{tabular}{lccc}
\hline
  Observation & Filter & Exposure time    & Limit \\
             Date       &        & (s)              & AB-mag \\
\hline
1998-11-11 & WFPC2/F336W & 2000 & 25.1 \\
1998-11-11 & WFPC2/F439W & 2000 & 24.3 \\
1998-11-11 & WFPC2/F555W & 660 & 24.5  \\
1998-11-11 & WFPC2/F814W & 720 & 23.8  \\
2004-10-23  &ACS/F555W & 480  &  25.2  \\
2004-10-23  &ACS/F814W & 720  &  24.6  \\
2004-11-01  &ACS/F435W    & 840           &   25.6 \\
2004-11-01   &ACS/F625W    & 360               &  25.5 \\
\hline
 \end{tabular}
\label{tab:obs2}
\end{table}

\section{Results}
\label{sect:results}

\begin{figure}
 \includegraphics[width=0.23\textwidth]{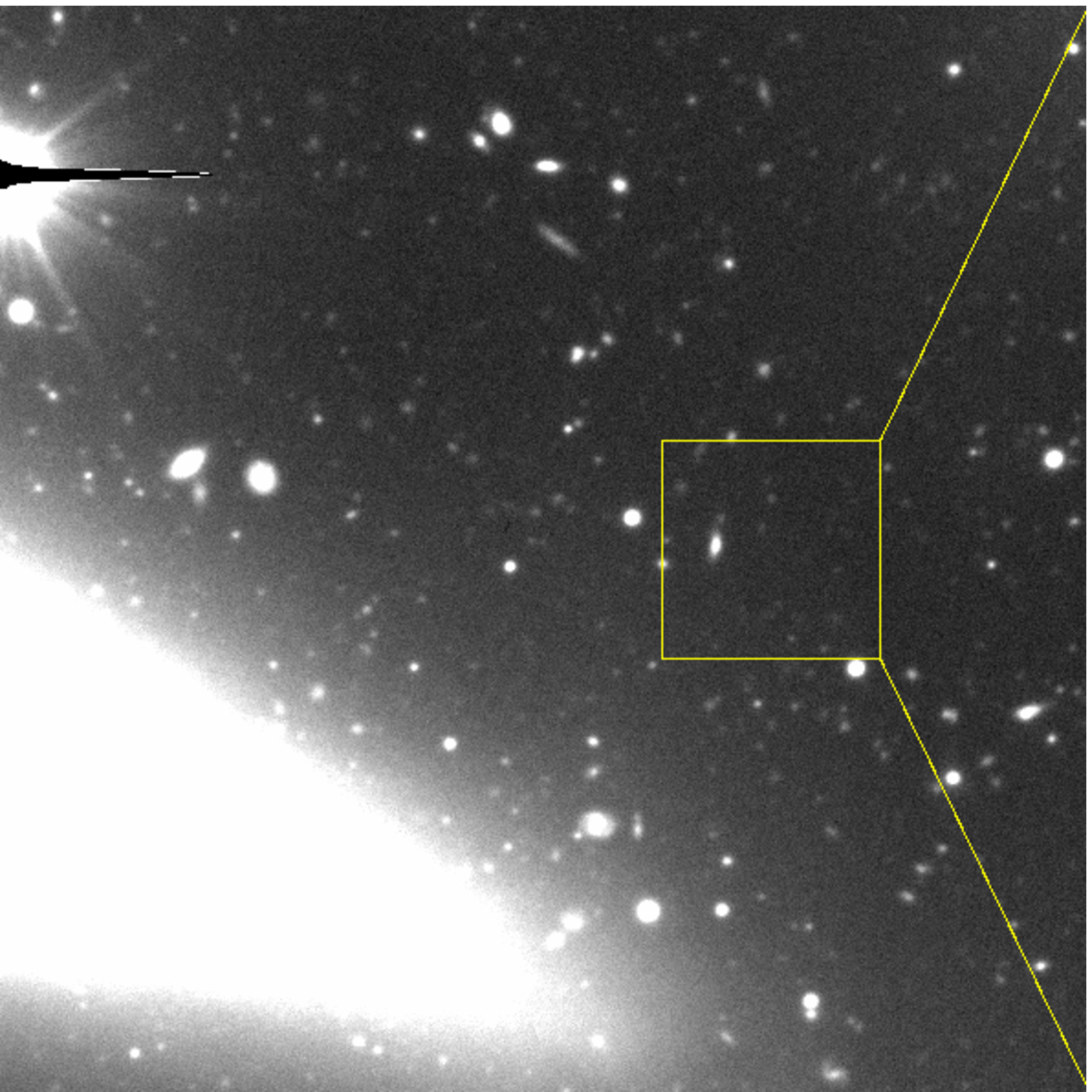}
 \hspace{-0.2cm}
 \includegraphics[width=0.23\textwidth]{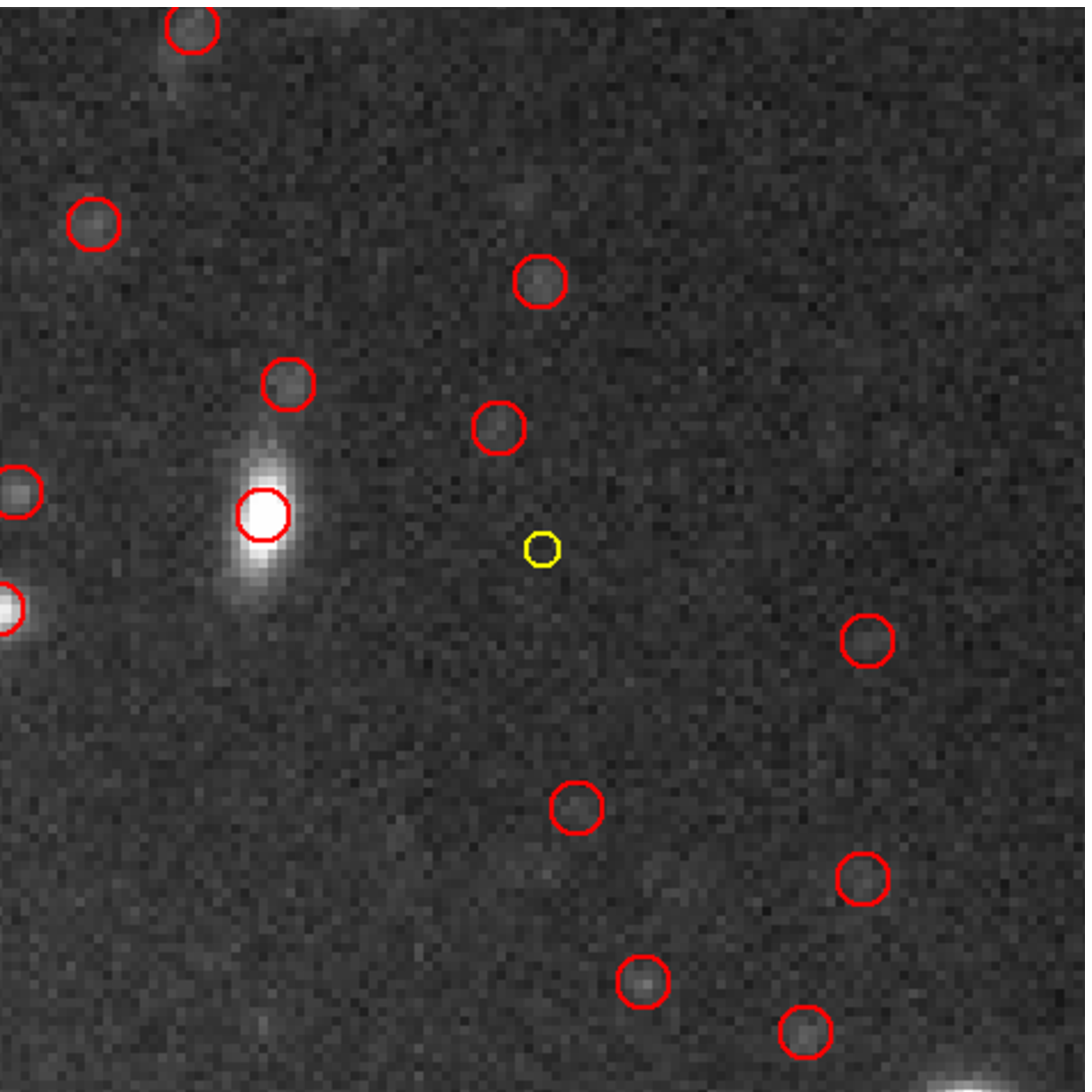}
 \caption{FORS2/VLT imaging of the location of SN~2005E in $R$-band. The transient location is marked in the zoomed-in section with a yellow 1 arcsec ($\sim$170~pc at the distance of NGC~1032) aperture. Objects detected in the images near the transient's location are indicated by red circles. The zoomed-in section is 30$\times$30~arcsec. North is up, east is left.}
 \label{fig:05E}
\end{figure}

\begin{figure*}
 \centering
 \includegraphics[clip=True,trim=0 0 0 1px,width=0.23\textwidth]{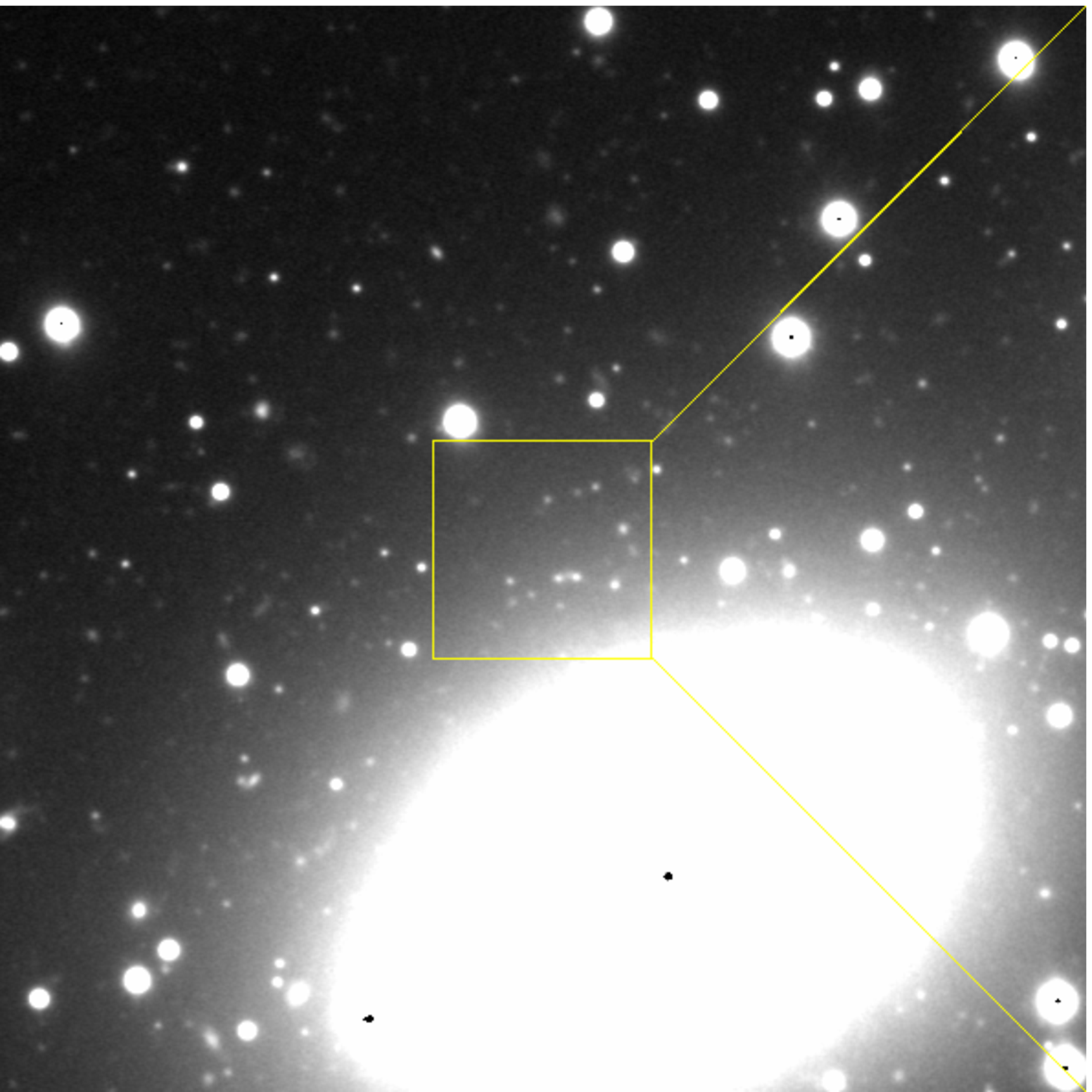}
 \hspace{-0.2cm}
 \includegraphics[width=0.23\textwidth]{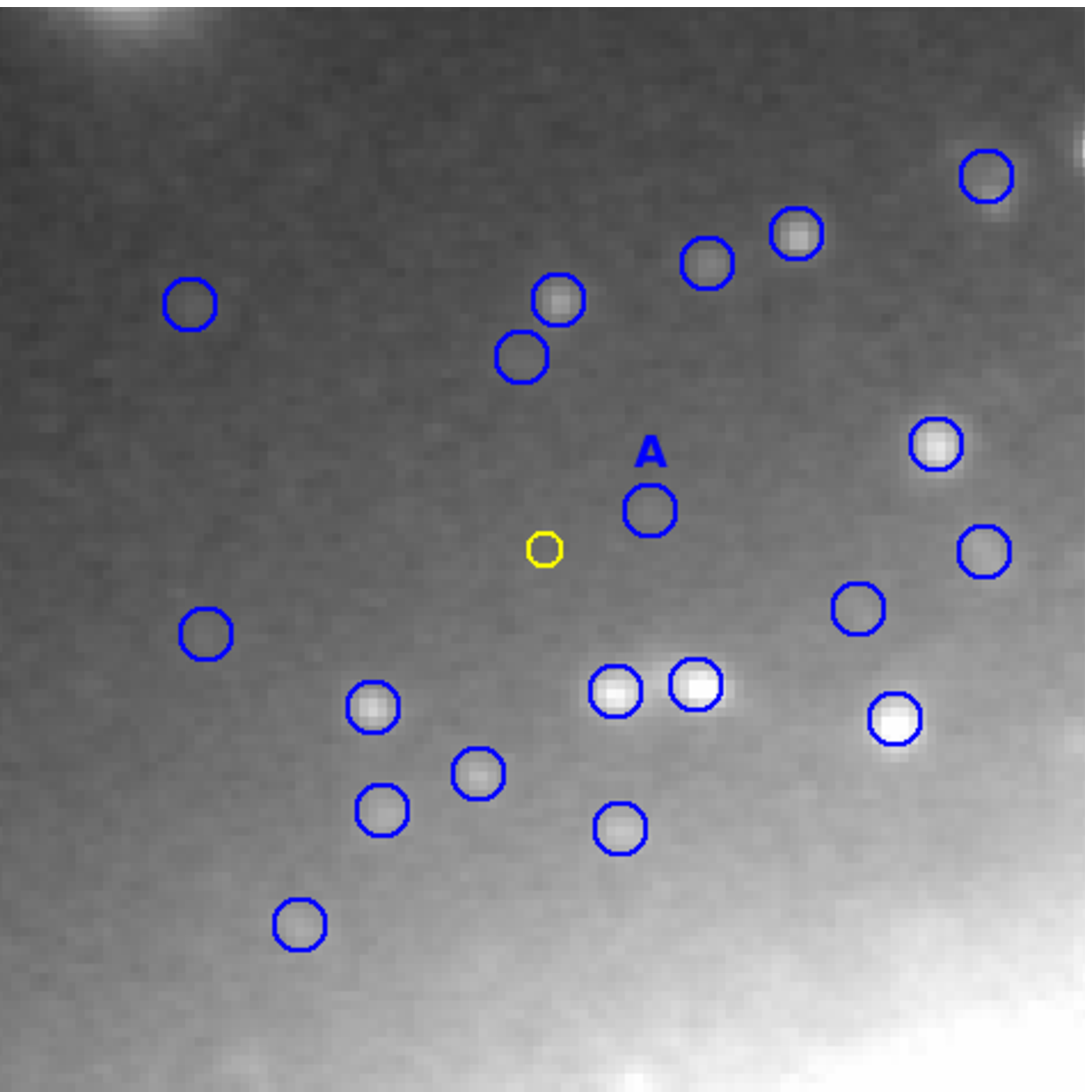}
 \hspace{0.5cm}
 \includegraphics[clip=True,trim=0 0 0 1px,width=0.23\textwidth]{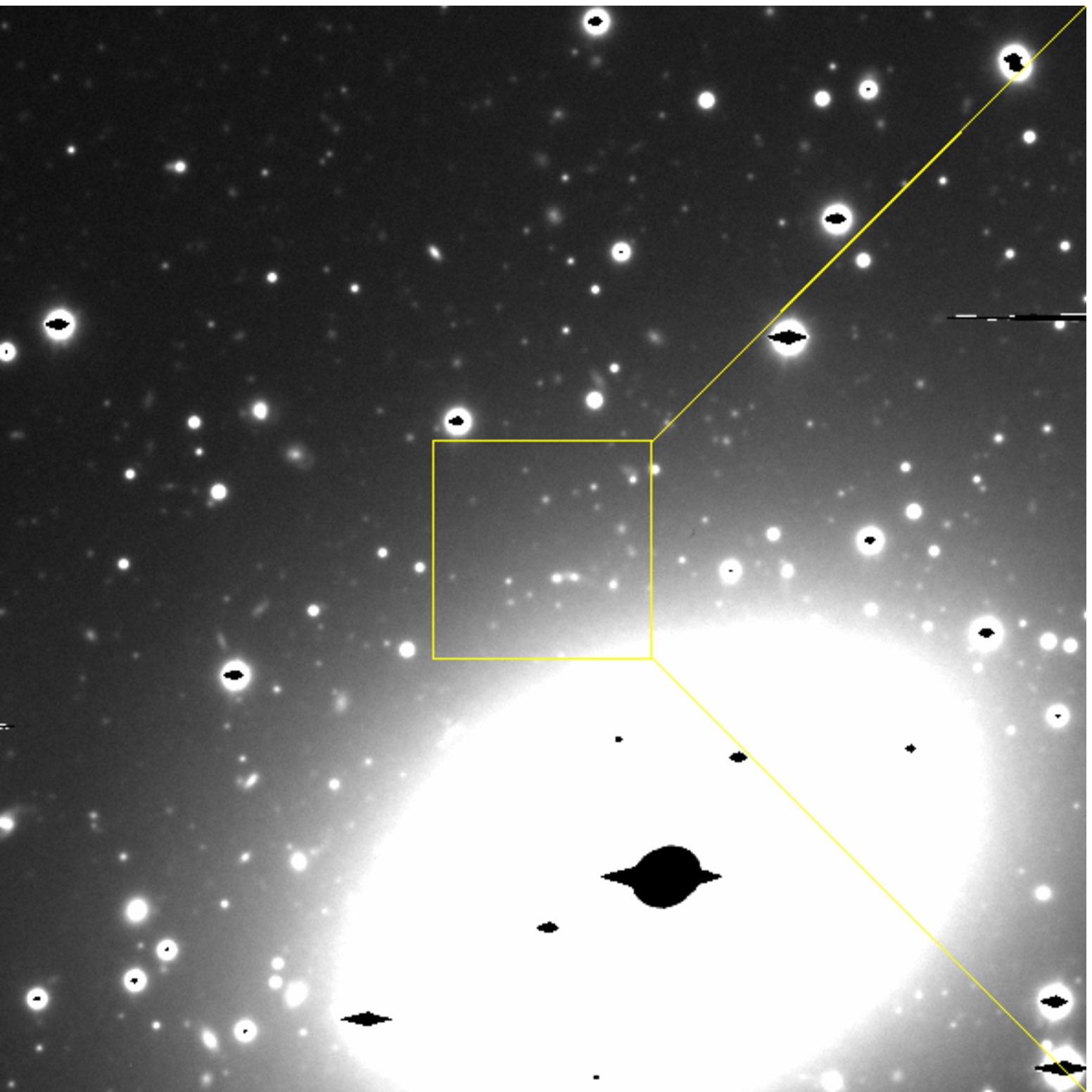}
 \hspace{-0.2cm}
 \includegraphics[width=0.23\textwidth]{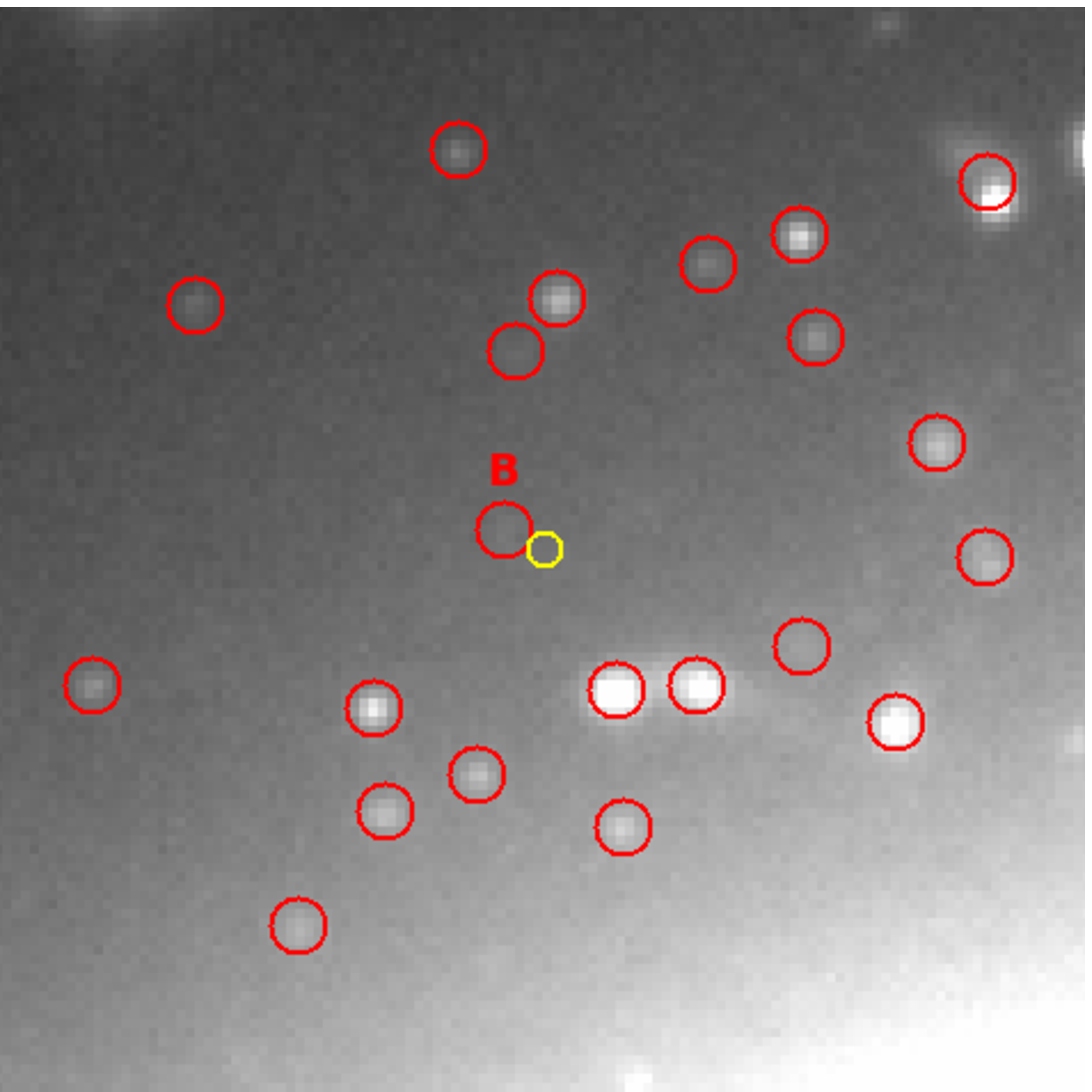}
 \caption{Same as \cref{fig:05E} but for SN~2012hn in $B$- and $R$-band (left and right, respectively). The 1 arcsec aperture is $\sim$150~pc at the distance of NGC~2272. The nearest object in each filter, as discussed in the text, is highlighted by an `A' and `B' label for $B$- and $R$-band respectively.}
 \label{fig:12hn}
\end{figure*}

\begin{figure}
 \centering
 \includegraphics[clip=True,trim=2.2cm 1cm 2.2cm 1cm,width=0.95\columnwidth]{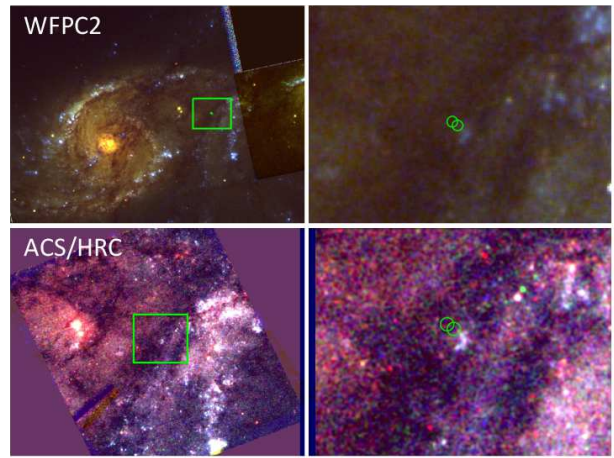}
 \caption{{\em HST} imaging of the location of SN~2003H. The top panels show WFPC2 observations taken in 1998, the lower panels ACS/HRC observations
 obtained in 2004 (composite images of filters given in \cref{tab:obs2}). Images are north up and east left. The right hand panels show zoomed in regions (corresponding to the green boxes on the left hand images). The
 two circles represent the position of the transient as determined from the host centroid and the offset star. The circles are 0.2 arcsec in radius. While the transient clearly lies on the stellar field of the merging host galaxy, close to the interface between the two, there is no obvious point source visible at the its location, although a faint star forming region is present offset approximately 1 arcsec to the SW of the transient's position (120 pc in projection). }
 \label{fig:03H}
\end{figure}

Images of the transients' locations from the FORS2/VLT, WFPC2/HST and ACS/HST imaging are shown in \cref{fig:05E,fig:12hn,fig:03H}. The location of SN~2012hn was found through star-matching alignment with archival imaging containing the luminous transient to obtain an accurate astrometric position in our imaging ($\sim$0.1--0.2~arcsec). For SN~2005E, imaging with the luminous transient detected was not available; the position used is that given by the WCS fit to the images. These WCS fits were found to be accurate when compared to catalogue star positions at \simlt{}arcsec accuracy. 
The location of SN~2003H was found by following offsets presented in \citet{graham03} from the neighbouring host and a nearby star. The two offsets gave a $0.2$~arcsec difference in the position. We adopt this as an uncertainty on our astrometry within the images and, where the uncertainty is significant, we include errors in photometric values by repeating measurements over this uncertainty and providing the more conservative estimate in each case. 

\subsection{Limits at location of transients}
\label{sect:tranloc}

Initially, aperture photometry was attempted at the locations SNe~2005E and 2012hn using 1 arcsec apertures.

No source was detected at the location of SN~2005E in our $R$-band imaging. A 3$\sigma$ magnitude limit was therefore determined at the location and then corrected from 1~arcsec to a large (4$\times$FWHM) aperture. Galactic extinction was accounted for \citep{schlafly11} with an $R=3.1$ extinction law, host reddening was assumed to be negligible given the 23~kpc offset of SN~2005E. The resulting intrinsic magnitude limit (i.e. after correcting for Galactic reddening) at the location was found to be $m_{R} > 27.4$~mag. Taking the distance modulus ($\mu$) to SN~2005E to be that of NGC~1032, $\mu = 32.74$~mag\footnote{Taken from NASA/IPAC Extragalactic Database (NED), assuming $H_0 = 70.4$~\kms{}~Mpc$^{-1}$ \citep{jarosik11}.}, we obtain an absolute magnitude limit of $M_{R} > -5.3$~mag.\footnote{This limit did not differ significantly when repeating measurements over the uncertainty on the position of SN~2005E.}

\begin{figure}
 \centering
 \includegraphics[width=0.23\textwidth]{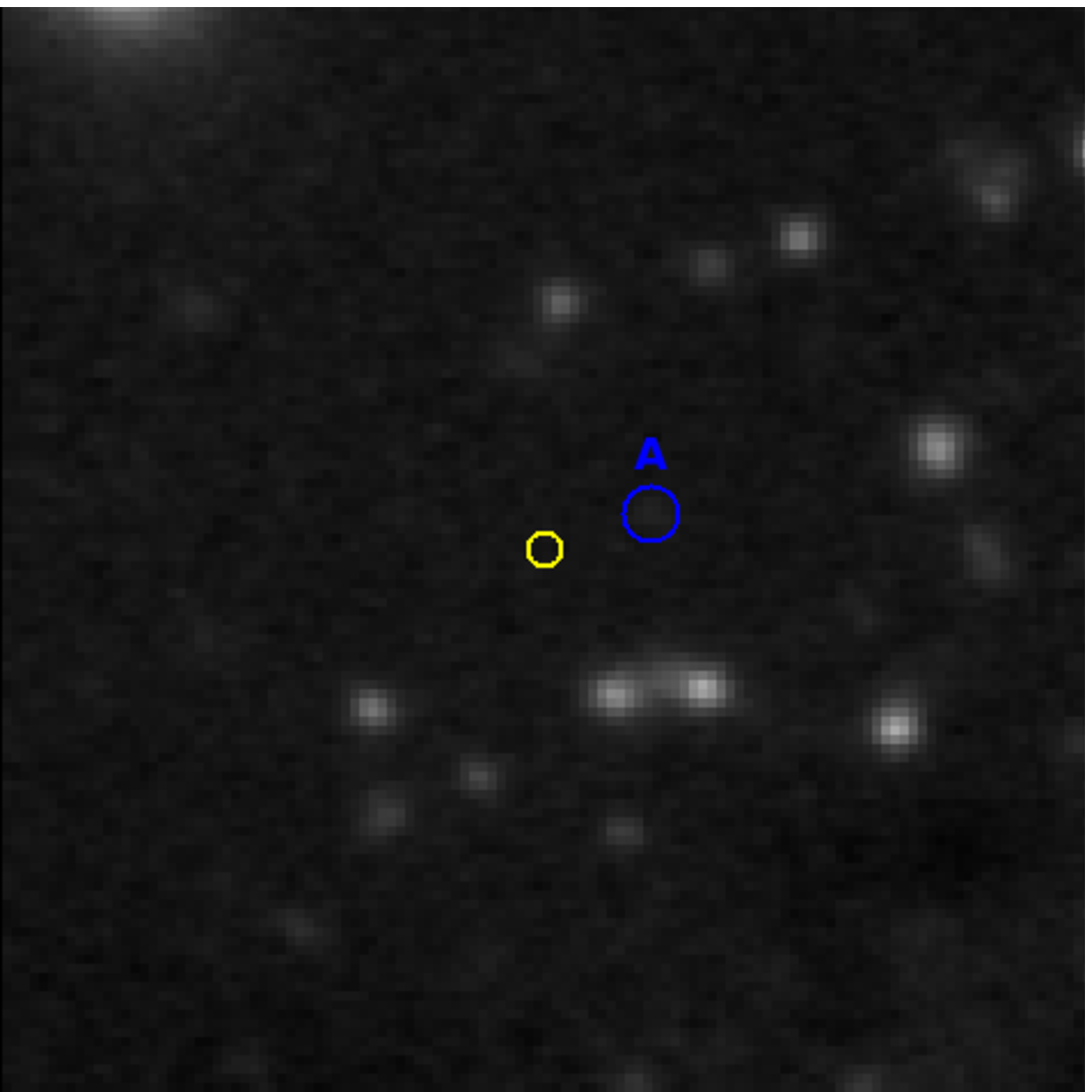}
 \includegraphics[width=0.23\textwidth]{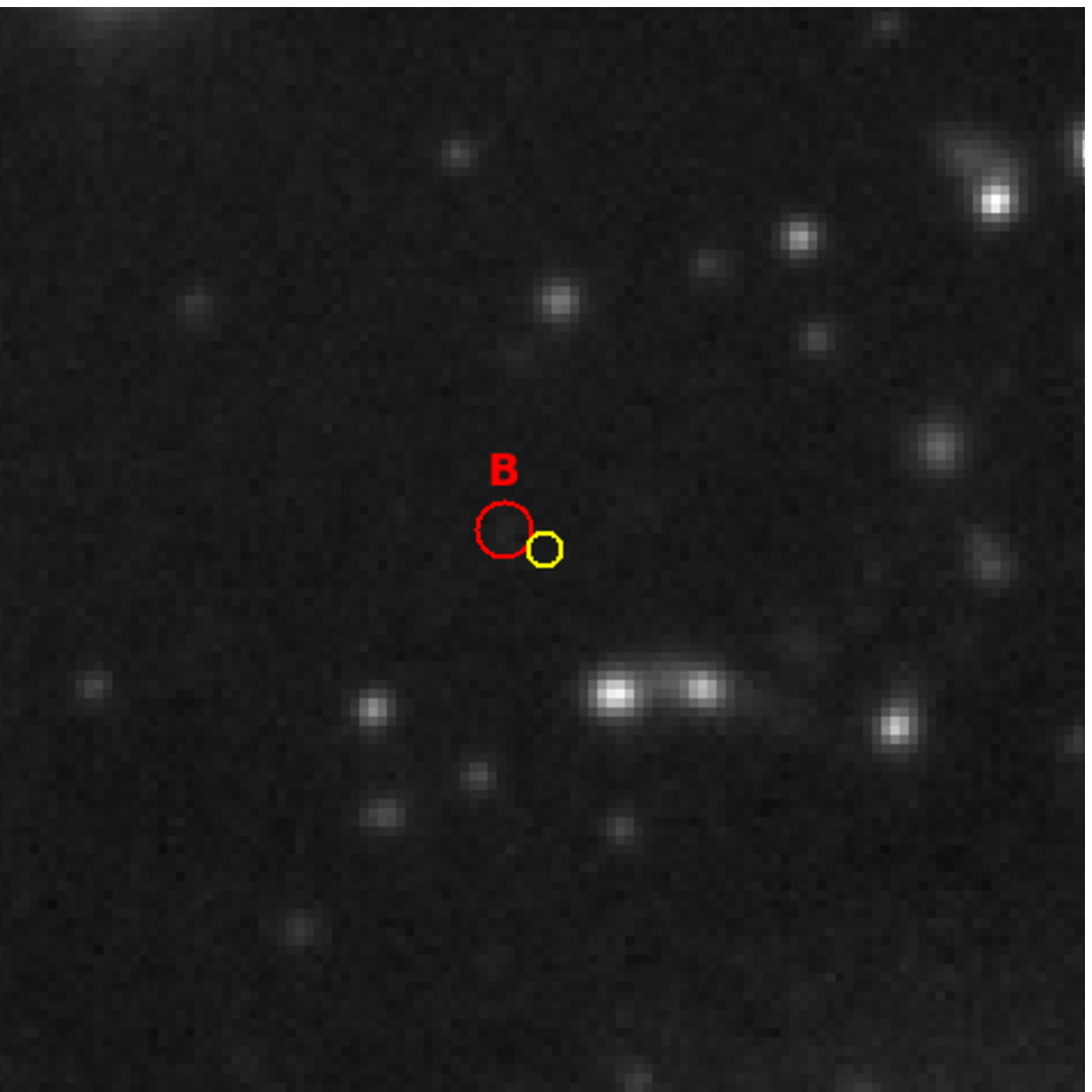}
 \caption{The location of SN~2012hn in $B$ (left) and $R$ (right) imaging after applying a smoothed spline to subtract the profile of the host-galaxy light. Pixel-value scaling was kept the same as the analogous images in \cref{fig:12hn}.}
 \label{fig:12hn_smooth}
\end{figure}

The location of SN~2012hn is situated in a region of significant surface brightness from the host, NGC~2272 (\cref{fig:12hn}). In order to subtract the shape of the galaxy profile in these outer regions, a function was fit to the background of the image. A two-iteration 3-sigma clip was applied to pixels across the image to remove objects but not affect the outer profile of the galaxy light (the central, very bright regions of the host were clipped, but cleanly subtracting the galaxy in these regions was not required). A median mesh with a box size of 25 pixels was then constructed and fitted with a smoothed spline. The spline function was subtracted from the original pixel values and the results are shown in \cref{fig:12hn_smooth}. Although over-subtraction arose near bright objects due to incomplete masking, the region of interest (i.e. the location of SN~2012hn) was found to have the large scale galaxy profile cleanly subtracted. As with SN~2005E, no source was detected in either our $B$- or $R$-band imaging at the location of SN~2012hn. Following the same procedure as for SN~2005E, from the background-subtracted images we find limiting magnitudes of any underlying object to be $m_{B} > 27.8$~mag and $m_{R} > 27.0$~mag. Taking $\mu = 32.56$~mag for NGC~2272 (from NED), these translate as absolute magnitude limits of $M_{B} > -4.8$~mag and $M_{R} > -5.6$~mag.

In contrast to the case of SN~2005E (which has essentially no underlying background galaxy) and SN~2012hn (which has a smoothly varying spheroidal component) the location of SN~2003H lies between the merging galaxies NGC 2207 and IC 2163. This region clearly contains underlying stars, and well as star formation, as is visible by large regions of blue star formation in the {\em HST} imaging. However, the astrometry shows that the transient lies away from the strongest regions of star formation, and is offset $\sim1$~arcsec from the nearest likely site of young massive stars. Indeed, within this merging region it appears to lie in a region of low stellar density, although this may be due to increased extinction along the line of site. We extract limits based on the measured flux at the location of SN~2003H, with the background calculated in apertures proximate to the position, but not overlying obvious structure. Given the high (and spatially variable) background these limits are significantly brighter than are in theory possible for the instrument. For a distance of 35 Mpc (\citealt{li11c}; $\mu = 32.72$~mag) the limits correspond to $M_B > -7.2$~mag and $M_R > -7.3$~mag.  Unlike other transients presented here, SN 2003H is unusual in having two epochs of observations, both before and after the explosion. We directly subtracted the images taken in the same (or similar) filters at different times by re-drizzling the ACS observations into the pixel scale and frame of those taken with WFPC2. No variable sources are found in the resulting frames, although the limits are weak: $m_{F439W} > 25.0 \pm 0.2$, $m_{F555W} > 25.5 \pm 0.4$, $m_{F814W} > 25.3 \pm 0.3$~mag, with the uncertainties arising from the uncertainty of the transient's position on the {\em HST} frames. Additionally, the location of SN~2003H is likely to be in a region of significant host reddening, and indeed spectra of SN~2003H are indicative of strong extinction \citep{IAUC8049}. The limits on the progenitor from the subtracted images do not provide tight constraints, ruling out only evolved stars $M_\text{ZAMS} > 15-20$~\msun{}, even when neglecting reddening (method given in \cref{sect:massive}). With a reasonable amount of reddening these limits become very unrestrictive. We therefore do not include the limits on SN~2003H in our subsequent analysis above general qualitative statements, given the intrinsic limits may be considerably shallower.

The limits given above are subject to the uncertainties in $\mu$ for each host ($\sim0.1-0.2$~mag), but these do not significantly affect the analysis.

\subsection{Nearby detections}
\label{sect:nearby}

{\sc SExtractor} \citep{bertin96} was used to search for detections local to the transients for SNe~2005E and 2012hn. The detections are shown in \cref{fig:05E,fig:12hn}. An object is detected north-east of SN~2005E, however this object is at a distance of $640$~pc in plane of sky alone ($\sim 3.8$~arcsec). The nearest detections to SN~2012hn are detected in only one band each. The blue object is located $\geq 460$~pc away and the red object is $\geq 190$~pc away, with magnitudes $m_B = 27.1\pm0.2$ and $m_R = 26.1\pm0.1$~mag respectively.

For SN~2003H the source clearly has stars underlying the position. However, it is not located within a region of massive star formation. The closest blue sources to its position 
lie approximately 1~arcsec away, 120 pc in projection at the distance of SN~2003H. The source has an AB magnitude of $m_{F435W} = 22.6 \pm 0.2$~mag, or an absolute magnitude of M$_B$ = -10.1~mag.

\section{Discussion}
\label{sect:discuss}

Here we discuss our results in the context of proposed progenitors or progenitor host systems.

\subsection{Massive stars}
\label{sect:massive}

\begin{figure}
 \includegraphics[width=\columnwidth]{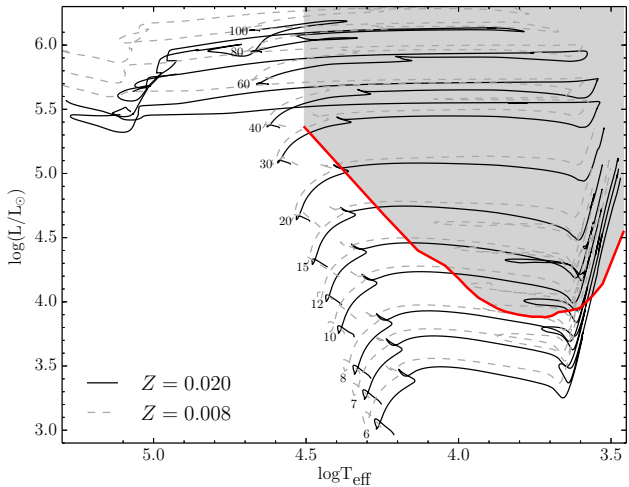}
 \caption{Detection limit at the location of SN~2005E in the HR diagram. The $R$-band limit is indicated by the red line. Overlaid are stellar models from BPASS for initial masses 6--100~\msun{} for metallicities $Z = 0.020$ and $0.008$.}
 \label{fig:05Ehr}
\end{figure}

\begin{figure}
 \includegraphics[width=\columnwidth]{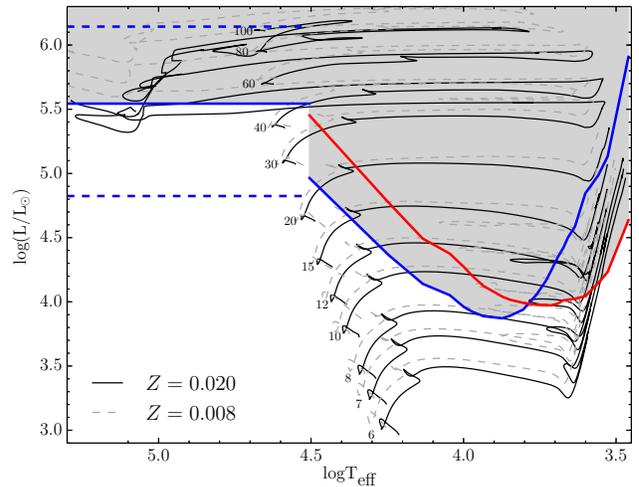}
 \caption{Detection limits at the location of SN~2012hn in the HR diagram. The $B$- and $R$-band limits are indicated by blue and red lines respectively. Overlaid are stellar models from BPASS for initial masses 6--100~\msun{} for metallicities $Z = 0.020$ and $0.008$. Also plotted is the WR luminosity limit for the $B$-band using the average BC to Galactic WR stars (solid blue horizontal line) and the limits on the BCs to WR stars found by \citet{crowther07} (dashed blue horizontal lines).}
 \label{fig:12hnhr}
\end{figure}

Although our VLT imaging is post-explosion, and as such we cannot observe the progenitor itself, it can still be used to shed light upon the possibility of a massive star origin for these transients. One method is to look for the surviving binary companion. Given the spectral signature of these transients was akin to a stripped-envelope SN, i.e. an absence of detected hydrogen, any massive star progenitor would need to have had its envelope removed throughout its lifetime. Recent studies suggest that stars in binary systems are the dominant progenitor channel for SNe exploding in this fashion (e.g. \citealt{eldridge08,dessart11,smith11b,lyman14c}; see also \citealt{smith14} and references therein), with the companion facilitating enhanced mass loss due to its gravitational presence \citep[e.g.][]{pod92}. This is consistent with massive star studies that highlight the prevalence for binarity in high mass stars \citep[e.g.][]{sana12}, and suggest that these companions are also massive \citep{kobulnicky07}. We can use our magnitude limits to constrain the mass of a surviving companion by comparing to stellar models in the H-R diagram. The methodology followed is similar to that in previous work \citep[e.g.][]{maund05,crockett07,perets11}, and is thus only briefly summarised here. 
Our magnitude limits were converted to limiting bolometric luminosity (\lbol{}) using the colours and bolometric corrections (BCs) for a range of supergiant spectral types provided by \citet{drilling00}. The \lbol{} limit as a function of effective temperature (\teff{}) could then be plotted in the H-R diagram (\cref{fig:05Ehr,fig:12hnhr}).
A Wolf-Rayet (WR) star limit was obtained in the manner of \citet{crockett07,perets11} -- the median WR colours of \citet{massey02} were used alongside the BC of \citet{smith89}, $-4.5$~mag, to obtain a limiting \lbol{} that was applied for high temperatures in the H-R diagram. The results of these limits are shown in \cref{fig:12hnhr} only, since the $R$-band limit for SN~2005E does not alone provide any stringent constraints on the WR regime. Note we also include the limits to the BC for WR stars found by \citet[][and reference therein]{crowther07} of $-6$ to $-2.7$~mag. These limits are compared to BPASS stellar models \citep{eldridge09}\footnote{\url{http://www.bpass.org.uk/}}, which are overlaid on our limits in \cref{fig:05Ehr,fig:12hnhr}.

We rule out the presence of all evolved stars $>10$~\msun{} and exclude the presence of a RSG at intermediate mass or greater at the location of SN~2005E. The lack of blue wavelength coverage means we do not have limits to impact on the regime of WR stars. Similar limits are imposed on the location of SN~2012hn, the additional $B$-band limits preclude the presence of younger moderately-massive stars. Taking the BC of \citet{smith89}, the $B$-band limit also limits any undetected WR star to be at the lowest mass range for WR stars at solar metallicity. Taking a lower metallicity, typical of the Large Magellanic Cloud, WR stars are further disfavoured; however, when considering the spread of possible BC for WR stars from \citet{crowther07}, we see that conclusively ruling out WR stars is not possible. The limits here, and the fact we are observing in broad-band filters as oppose to the narrow-band filters typically used for WR observation (which means the limiting \Mbol{} may be up to $\sim$0.5~mag fainter, see discussion in \citealt{crockett07}) make the presence of a WR disfavoured, but not conclusively excluded. The limits from the post-explosion {\em HST} observations of SN~2003H rule out the presence of underlying very massive stars, however, are not restrictive enough to put limits on more moderately massive stars.

The above discussion thus strongly limits the possibility for the progenitor to have had a massive star companion. Even neglecting the need for a binary companion, massive stars form in associations \citep{lada03} and thus other massive stars will have formed contemporaneously at the same location. By ruling out the presence of any other massive stars to stringent limits, we can infer the progenitors were not massive stars since we would expect to see the other massive cluster-members, or clusters nearby that may have formed the progenitors. 

For SN~2000ds, \citet{maund05} exclude a RSG progenitor with $M_\text{ZAMS}$~\simgt{}~$7$~\msun{} from pre-explosion imaging, but cannot rule out a blue progenitor. Similarly, \citet{perets11} present pre-explosion imaging of SN~2005cz that rules out the progenitor being an evolved star with $M_\text{ZAMS}$~\simgt{}~$15$~\msun{}. Additional post-explosion imaging restricts the presence of RSGs to those with $M_\text{ZAMS}$~\simlt{}~$10$~\msun{}, strongly disfavouring recent star formation at the location, or the presence of a binary companion to SN~2005cz.

Hypervelocity massive stars can also be ruled out as a means to alleviate the requirement for underlying star formation. Notwithstanding the complete lack of recent star formation seen in half of the hosts of \cas{} \citep{lyman13}, the short life times of massive stars means it is difficult for such a star to survive out to tens of kpc from their host (in plane of sky alone) before core-collapse without introducing unreasonably high velocities. This has been noted previously, with both \citet{perets10} and \citet{kasliwal12} arguing against the case of massive runaways.

\subsection{Dwarf galaxies}
\label{sect:dwarfgal}

\begin{figure}
 \includegraphics[width=\columnwidth]{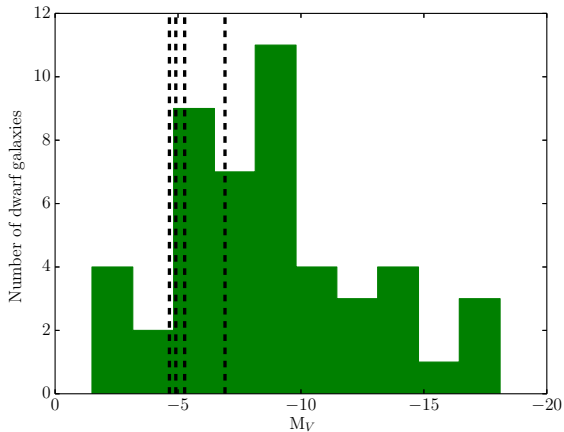}
 \caption{Detection limits at the locations of four Ca-rich transients, SNe 2000ds (M$_V = -4.7$), 2005E (M$_V = -4.9$), 2005cz (M$_V = -6.9$) and 2012hn (M$_V = -5.3$), are shown compared with the M$_V$ of local group dwarf satellites \citep{mcconnachie12}. The limits for 2005E and 2012hn were converted using the median colours of dwarf galaxies from \citet{mateo98}, see text.}
 \label{fig:dwarfgal}
\end{figure}

Dwarf galaxies may explain the large offsets exhibited by some \cas{}, and may be a favoured model in the case that very low metallicity is required for their
production. The locations of PTF~09dav, 10iuv (= SN~2010et) and 11bij have limits of underlying dwarf hosts to be M$_R > -9.8, -12.1$ and $-12.4$~mag respectively \citep{kasliwal12}. The PTF-discovered examples are at much farther distances than those discussed here, although the limits do preclude brighter satellite hosts underlying their locations. \citet{mcconnachie12} present data for dwarf satellites for local galaxies, including M$_V$ values. The median $V-R$ colour of the dwarf galaxy sample of \citet{mateo98}, $V-R = 0.50\pm0.17$, was used to make an estimate of our $R$ limits in the $V$-band for dwarf galaxies. The distribution of dwarf galaxy luminosity is presented in \cref{fig:dwarfgal} alongside our limits on 2005E and 2012hn and literature limits for 2000ds \citep{maund05} and 2005cz \citep{perets11}. These literature limits were converted from the photometric system of {\em HST} to Johnson $V$-band using the method of \citet{sirianni05}, using $B-V = 0.71$, the median of dwarf satellites presented in \citet{mateo98,kim11}.

At first glance, it appears that any potential faint systems we have not detected are not being drawn from the overall population -- the two-sample Kolmogorov–Smirnov test gives a $2\sigma$ rejection, this is a lower bound since we only utilise limits, however there are important caveats to consider. Due to the difficulty in finding extremely faint systems, it is almost certainly the case that the left side of the galaxy distribution in \cref{fig:dwarfgal} is deficient due to observational bias (further exacerbated by the lack of adequate coverage and depth over large portions of the sky). The prevalence of ultra-faint systems is currently very uncertain and poses a quandary between observation and theory, however, is it likely that the stellar budget contributed by these extremely dark-matter-dominated systems is much less that those of brighter satellites. With this in mind, one would expect to observe \cas{} exploding in moderate-to-luminous dwarf galaxies, making the reasonable assumption that the chance for a progenitor to form scales with the luminosity (stellar mass) of a galaxy. Unless one introduces an unreasonably large number of missed ultra-faint systems, or there exist hitherto unknown environmental factors that make ultra-faint satellites preferential producers of Ca-rich transients, a dwarf satellite origin is disfavoured by the limits present on these nearest Ca-rich systems.

\subsection{Globular clusters}
\label{sect:gc}

Globular clusters represent a promising birth site for unusual transients at large galactic radii (i.e. in the halo) since their dense stellar cores enable dynamical interactions to create and harden exotic binaries, leading to greatly enhanced production of such systems \citep[e.g.][]{davies95,davies95b,pooley03,ivanova06,grindlay06}. Hence it is relevant to search for globular clusters directly under Ca-rich transient positions. At these distances they should appear as point sources to ground based observations (0.7~arcsec seeing at a distance of 30~Mpc gives a resolution of $\sim$100~pc, much larger than GC scale sizes), and barely resolved in {\em HST} data. 

Encounter rates in clusters will generally increase with cluster mass (see for example equation 1 from \citealt{davies04}). Indeed fewer than ten globular clusters account for over half of all dynamical encounters occurring within the entire globular cluster population of our Galaxy \citep[e.g.][their figure 1]{verbunt87}. For example, the cluster 47 Tuc, accounts for some 10 per cent of all encounters.

The globular cluster luminosity function (GCLF) of the Milky Way was constructed using the catalogue of \citet{harris96} (2010 edition).\footnote{\url{http://physwww.mcmaster.ca/~harris/mwgc.dat}} Since the GCLF is well described by a normal distribution \citep{harris91}, we can ascertain the probability of not detecting a faint, underlying GC by fitting a normal distribution to the GCLF and extracting the value of the cumulative probability density function at our limits. This was performed on the raw GCLF and also on the luminosity-weighted GCLF, since the ability for a GC to produce a transient scales with the cluster mass, and thus luminosity.

\begin{figure}
 \includegraphics[width=\columnwidth]{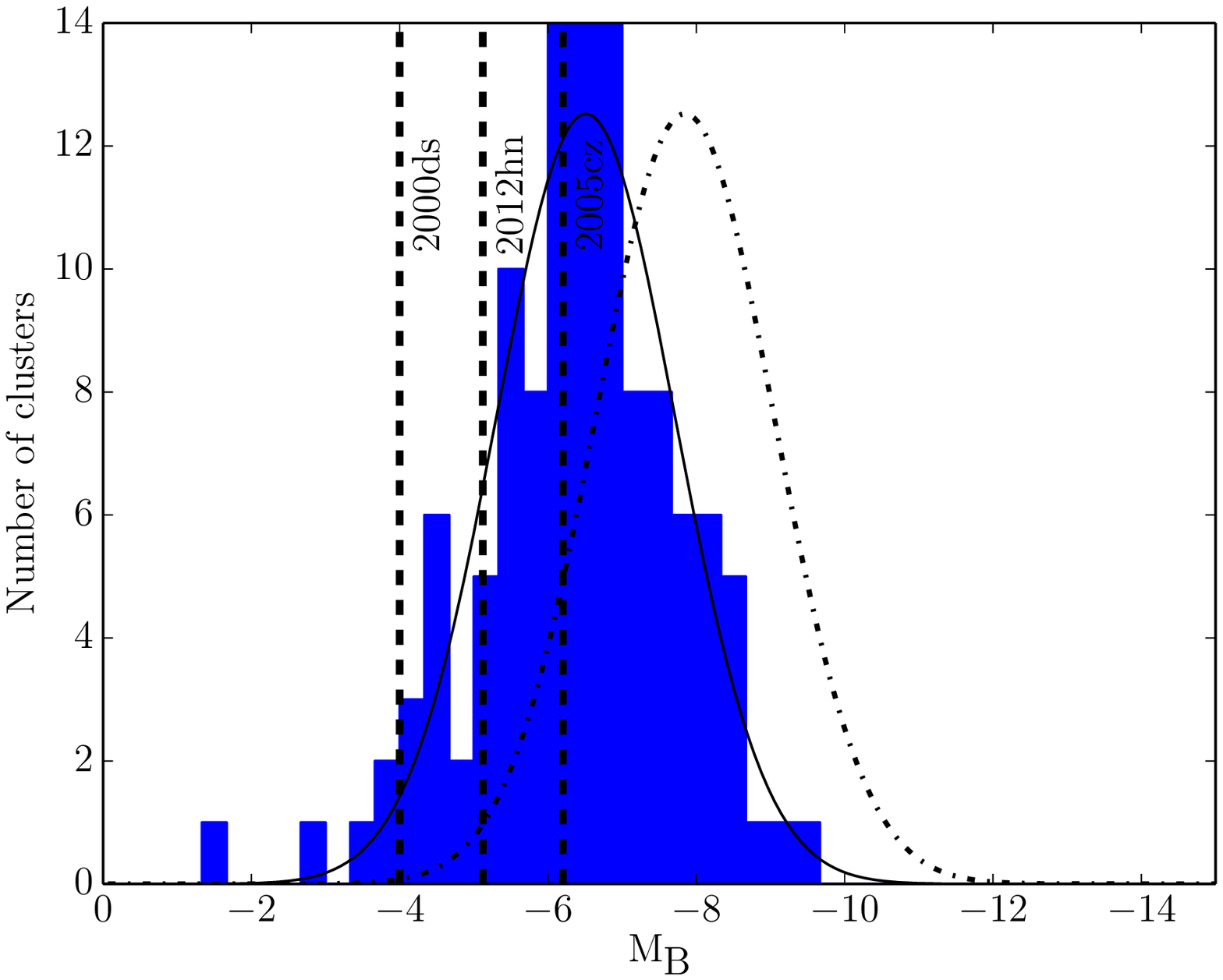}\\
 \includegraphics[width=\columnwidth]{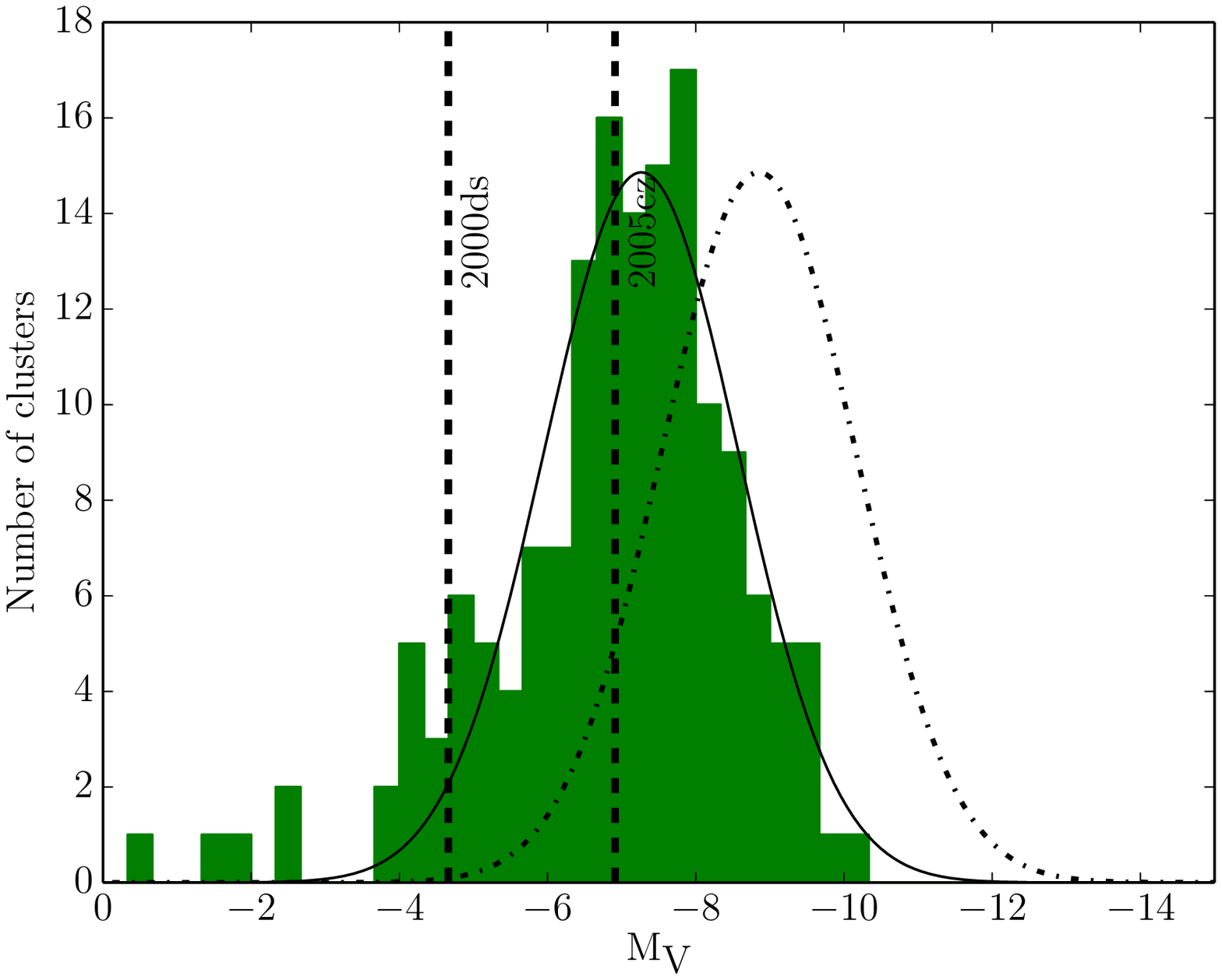}\\
 \includegraphics[width=\columnwidth]{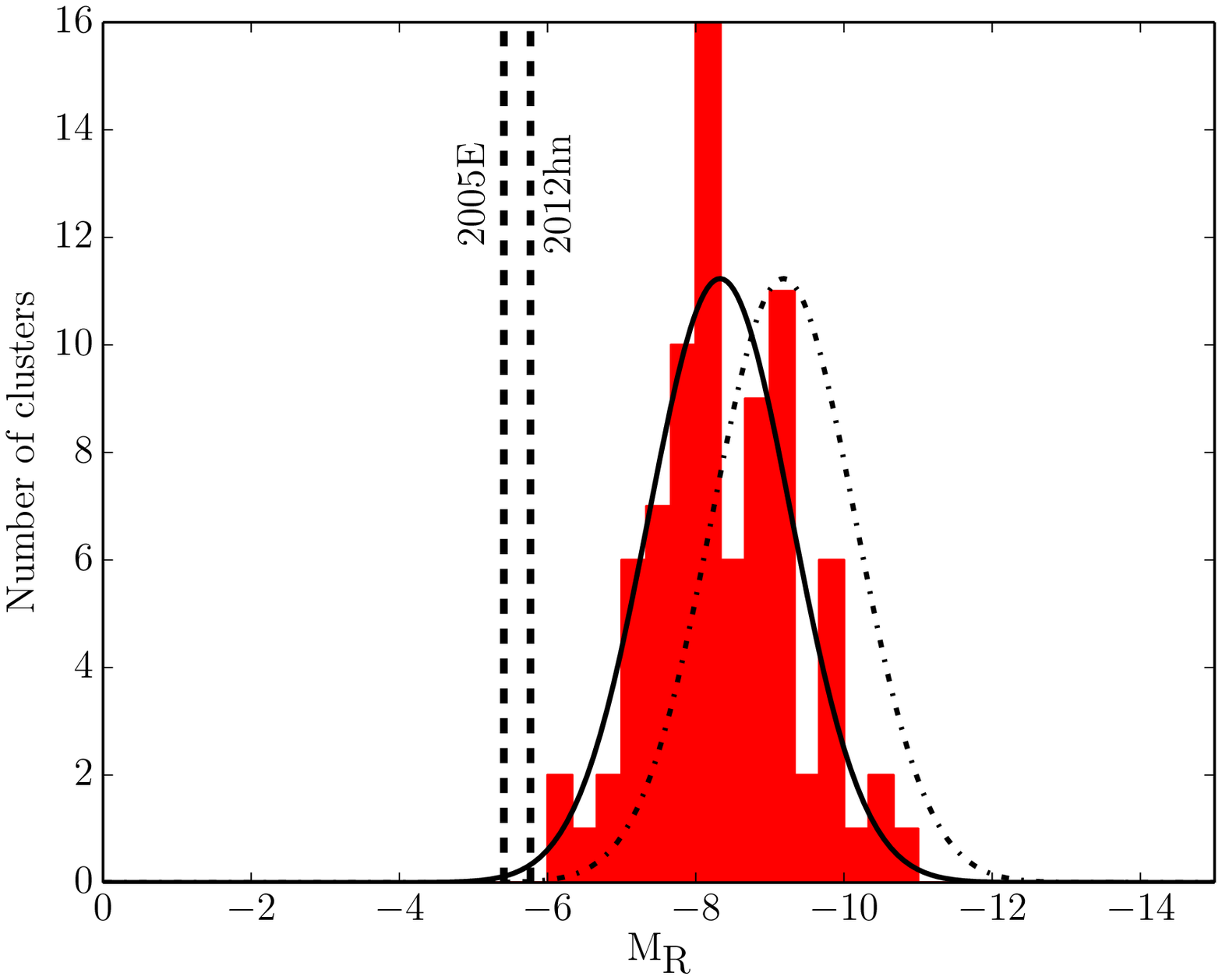}
 \caption{The GCLF from \citet{harris96} (2010 edition). For each filter, a normal distribution is fitted (solid line). Magnitude limits for the Ca-rich transients analysed here are shown as vertical dashed lines, also included are the post-explosion limits of SNe~2000ds \citep{maund05} and 2005cz \citep{perets11}. The luminosity-weighted GCLF is shown as a dot-dashed line in each plot.} 
 \label{fig:gclf}
\end{figure}

GCLFs for $B$- $V$- and $R$-band are shown in \cref{fig:gclf}, alongside limits from our observations. Also displayed are literature limits for SNe~2000ds and 2005cz from \citet{maund05} and \citet{perets11}, respectively. As in \cref{sect:dwarfgal}, these have been converted using the method presented in \citet{sirianni05}, using the median colour of globular clusters from \citet{harris96} (2010 edition) -- $B-V = 0.69$. It is clear the deepest of these limits are probing well down to the faintest GCs. Taking the combined probabilities for all transients in each filter, we can rule out the non-detections being due to underlying GCs on the very faint tail of the GCLF at $\sim$~3, 2.5 and 4~$\sigma$ for $B$- $V$- and $R$-bands respectively. When weighting by luminosity, following the prescription in \citet{harris91}, an underlying, undetected GC population is ruled out at $4-6\sigma$ for each band. The limits on the PTF-discovered events by \citet{kasliwal12} are not deep enough to increase these probabilities noticeably.

The nearest detected object to SN~2012hn, which is consistent with a GC at $B-R > 1.7$, is at a distance of $\geq 190$~pc. Such a large offset is difficult to reconcile with the more modest radii of GCs. Additionally, when considering the more modest host offset of SN~2012hn and the Galactocentric distance to half-light radius ($r_{h}$) relation for GCs \citep{vandenbergh94}, one may expect local GCs to be comparatively compact. For the majority of GCs, $r_{h} < 10$~pc, with a few per~cent showing $10 \leq r_{h} \leq 27$~pc \citep{harris96} (2010 edition). \citet{mackey06} present data for extended GCs, located at extreme offsets to M31, however even these unusual clusters exhibit $r_{h}$~\simlt{}~$35$~pc. As such, membership of SN~2012hn to this system (if indeed it is a GC) is strongly disfavoured.

\subsection{Compact binaries}

Interactions or mergers within binary systems of various configurations of degenerate bodies represent a potential progenitor channel for \cas{} \citep[e.g.][]{bildsten07,perets10,brown11,waldman11,metzger12}. Due to the dim nature of the component bodies, such binaries would be much fainter than our detection limits, being even difficult to observe within the Galaxy. As such, we cannot put meaningful restrictions on the composition of potential compact binary systems directly from our observational limits, but one can assess the potential for various configurations to explain the locations of \cas{} given we have found no convincing host systems underlying the locations.
Proposed progenitors must explain the lack of association that \cas{} have with their hosts stellar light, and specifically explain the remote locations that contain no underlying host systems, as shown by previous sections.

Although GCs do not appear to underlie the \cas{} analysed here, the possibility that at least a fraction of the progenitors are halo objects, formed at these large offsets, remains. In this case WD-WD binaries represent a potential progenitor channel.
WD-WD systems containing an extremely low mass WD (\simlt{}~0.25~\msun{}) are seen in the thick disc and halo of the Galaxy \citep{brown10}, however, detecting them beyond a few kpc is extremely difficult, and thus a census of the locations of such systems in and around the Galaxy is lacking. The mergers of these systems have been analysed as progenitors of subluminous SNe \citep{brown11} as part of the general `double-degenerate' scenario of WD-WD binary progenitors.
A significant fraction of SNe~Ia are also thought to arise from WD-WD binaries \citep[see recent reviews of][]{maoz13,ruiz14}, yet SNe~Ia are seen to follow the stellar light of their hosts \citep[e.g.][]{forster08, kelly08}. Although there may be contamination from other progenitor channels, SNe~Ia are rarely seen at very remote locations. \citet{kasliwal12} find only a few per~cent of SNe~Ia at offsets of $>$~30~kpc from a PTF-selected sample, in contrast to 3/3 \cas{} from their sample.\footnote{SNe~Ia have been seen to explode in low-luminosity hosts \citep[e.g.][]{strolger02}, which may not have been be detected in the survey or follow up imaging -- this may provide an explanation for the very remote SNe~Ia. The limits we present probe much deeper than the luminosities of such hosts.} The fact SNe~Ia follow the stellar light of their hosts, with such a low fraction of remote events, is consistent with expectations that WD-WD binaries trace the mass of their host galaxies, consequently we would expect \cas{} to also follow the stellar light of their hosts and have a low fraction of remote events, if arising from WD-WD binaries.

\begin{figure}
 \includegraphics[width=\columnwidth]{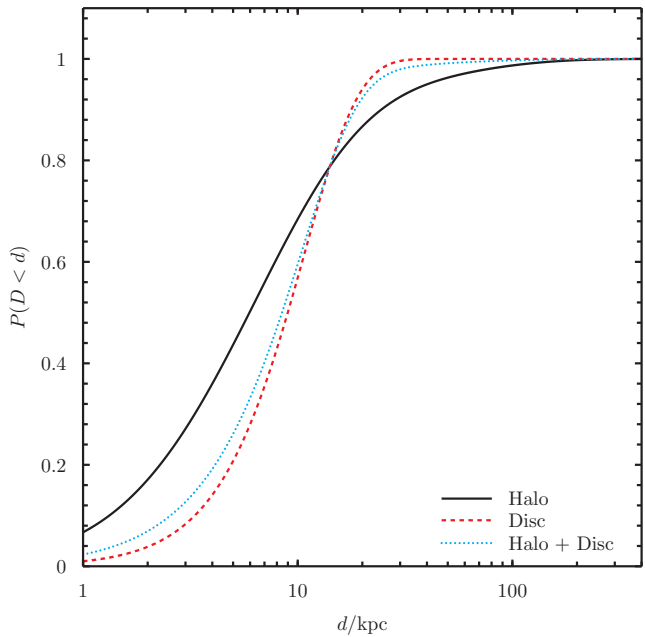}
 \caption{The fraction of disc and halo light interior to a galactocentric distance, $d$ for M31. Although extended, the halo remains centrally concentrated. Only a small fraction of the halo light extended to the distances of remote \cas{}, see text.} 
 \label{fig:m31}
\end{figure}

One may introduce the necessity for extremely old progenitor systems or low metallicity to make the halo a preferential site of their production in order to overcome a correlation with stellar light. However, halo profiles of galaxies, although extended, are nevertheless centrally concentrated. \cref{fig:m31} shows the radial profiles of the disc and halo components of M31 from the fits of \citet{tamm07}. Even for the very large M31, 90~per~cent of the halo lies within 20~kpc of the galactic centre, slightly larger than the disc extent \citep[see also, e.g.,][]{courteau11}. Thus, even if \cas{} were {\em solely} comprised of halo objects, we would still only expect to observe around 10~per~cent of \cas{} at \simgt{}~20~kpc offsets (cf. one third observed). When coupled with studies suggesting a large, likely dominant, proportion of Galactic WD-WD systems are formed and reside in the disc \citep[e.g.][]{bergeron03,kawka04,brown10}, further reducing the fraction of systems $>$~20~kpc (\cref{fig:m31}), it appears the locations of \cas{} do not agree with the expected locations of WD-WD binaries. This will be further studied in \citet{churchprep}.

\subsubsection{Kicked systems}
\label{sect:kicked}

The remote locations of these unusual transients may be indicative of high velocity, or runaway, systems. This possibility becomes favourable given the lack of detections of underlying systems at their locations, and that they do not follow the expectations of even a halo origin. Two main mechanisms can impart a high velocity on a system -- either through dynamical interaction \citep[e.g.][]{gvaramadze09}, or through a SN kick in a binary (due both to mass loss and natal kicks; e.g. \citealt{church11}).

Since hypervelocity massive stars cannot explain the locations of \cas{} (\cref{sect:massive}), one can infer that any high velocity progenitors would have to be lower mass (and thus longer lived) and also in binary systems, since there would be no (known) mechanism for a single, low mass object to undergo explosion if unmolested. Compact degenerate binary systems having undergone a kick then become a more favourable prospect.

The contribution of kicked-systems to the progenitors of CCSNe and gamma-ray bursts (GRBs) has been analysed previously \citep[e.g.][]{eldridge11,fong13}. In particular there are some parallels between \cas{} and short-duration GRBs (SGRBs) when considering their host environments, which are not true of the environments of other extragalactic transients. 

SGRBs do not follow the optical or UV light of their hosts \citep{fong10,fong13b}, similar to \cas{}. This is not the case for long-duration GRBs, core-collapse and thermonuclear SNe \citep{fruchter06,forster08,kelly08,anderson12} and the peculiar SN~2002cx-like transients \citep{lyman14b}, which all display a strong correlation to optical and/or UV light. About 15-30~per~cent of SGRBs appear `hostless' \citep{fong13,tunnicliffe14}, i.e. there is no detected underlying host galaxy, although their proximity to a relatively nearby galaxy is above that of random locations on the sky \citep{tunnicliffe14}, suggesting they have been kicked from these galaxies. Offsets from putative nearby hosts reveal a quarter of SGRBs have offsets of $>$~10~kpc, with around 10~per~cent at $>$~20~kpc \citep{fong13b}. The median SGRB offset is $\sim$5~kpc \citep{berger13}, larger than that of core-collapse and thermonuclear SNe, and particularly of the more centrally concentrated stripped-envelope CCSNe \citep{prieto08,anderson09}, which \cas{} most closely resemble spectrally. For \cas{} the median offset is $\sim$7~kpc (\cref{tab:offsets}, when considering the nearest putative host). The hosts of SGRBs are diverse, with a large fraction of non-star-forming hosts -- $\sim$20--40~per~cent being early-type \citep{fong13}, for \cas{} this is 50~per~cent \citep{perets10,lyman13}. These unusual characteristics of SGRBs have been used to argue in favour of kicked systems being the progenitors, which may provide an explanation for the characteristics of \cas{} also.

In the case of SGRBs it is believed that the progenitors are NS-NS or NS-BH binaries, with the recent detection of a kilonova offering evidence of this model \citep{tanvir13}. These systems
are kicked due to supernovae and neutron star natal kicks, and the locations of the SGRBs around the hosts is broadly consistent with the expectations of these models
\citep{church11,berger13}. For \cas{} it seems unlikely that such progenitors should produce these events, although their peak magnitudes and evolution timescales
are similar to those predicted for kilonovae \citep{metzger12b}. An alternative is that they arise from WD-NS mergers, which can be produced via similar routes to those
postulated for NS-NS and NS-BH systems \citep{davies02,church06}, and indeed these have been suggested as progenitors for \cas{} \citep{metzger12}. In a forthcoming
paper \citep{churchprep} we will consider whether the observed locations are consistent with the expectations of WD-NS models. It has been suggested that WD-NS mergers could produce long-duration GRB-like transients \citep{king07}. Deep observations in the high-energy regime would then be motivated for future \cas{}.

\subsection{Sample definition}
\label{sect:sampledef}

A cautionary note must be made since we are analysing a new and emerging class of transient. As such, membership of the class may be contaminated by intrinsically different transients (this is particularly true when reanalysing historical SNe that may not have complete photometric coverage). \citet{kasliwal12} consider 5 events (PTF~09dav, 10iuv, 11bij and SNe~2005E, 2007ke ) to satisfy all of the criteria they establish for the class. Subsequently, the well observed SN~2012hn \citet{valenti14} became another addition. Even if we take this strict cut on membership, we can still utilise our limits of SNe~2005E and 2012hn to show there are no underlying GCs (\cref{fig:gclf}) or underlying recent star formation, as traced by massive stars (\cref{fig:05Ehr,fig:12hnhr}) at the locations of the two most proximate examples, where such deep limits can be placed. The presence of underlying dwarf galaxies would still be disfavoured (when coupled with the limits for three PTF-discovered members presented by \citealt{kasliwal12}). Additionally, imposing this strict membership requirement would only serve to enhance the extremity of the environments and galactocentric offsets of the \cas{}. Using the offsets from \citet{kasliwal12} and \citep{valenti14}, the median (mean) offset for the sample of 6 events is 28 (25)~kpc -- further strengthening a lack of association between the locations of \cas{} and the stellar mass of their hosts.

\section{Conclusions}
\label{sect:conc}

We have presented deep observations of a sample of very local \simlt{}35~Mpc \cas{}, using recent VLT and archival {\em HST} observations, and pre-existing literature limits on other members. 
These observations yield no clear detections, but provide the strongest constraints on the progenitors and the host systems thus far. Limiting magnitudes are fainter than massive ($>10$~\msun{}) stars and young star clusters, ruling out the presence of recent star formation and thus further disfavouring a massive star origin for \cas{}, in line with previous work on direct detection of the progenitors and host environments. Faint dwarf satellites hosts are strongly disfavoured from a comparison to the luminosity of satellites around local galaxies. 
Underlying globular clusters are ruled out from comparing the limiting magnitudes to the distribution of Galactic systems. Given a lack of detected systems, the extremely low stellar densities at the extreme galactocentric offsets a substantial fraction of \cas{} exhibit (4/12 at $>$~20~kpc in projection alone) means their production would have to be enormously enhanced in the physical conditions of these locations in order to avoid a correlation with the stellar light of their hosts.

Such remote locations (where there exist no convincing underlying systems) pose a problem for most progenitor system models, which would be expected to have a correlation with the stellar mass (light) of their hosts.
The remote locations are difficult to reconcile with even an entirely halo-borne progenitor population.
An interpretation to explain the remote locations is that the progenitors of \cas{} are not born at the sites where they explode.  Hence, models in which the progenitors have been kicked from their birth sites, most likely due to supernova kicks, would be favoured, alongside a delay time distribution before explosion extending to long time-scales to explain offsets of tens of kpcs. In this case, WD-NS mergers may represent a promising progenitor channel, although further studies will be required to attempt to reproduce the spectral and photometric characteristics of these transients. Given the similarities in the evolution routes to produce NS-NS and WD-NS binary systems, the Ca-rich transients would then provide new constraints on the likely creation rate of all double compact object binaries, and have broad implications for both near term high-frequency, and long term low frequency gravitational wave searches.

\section{Acknowledgements}

We thank the referee for their comments, which improved the clarity of the paper. Danny Steeghs and J.~J. Hermes are thanked for useful discussions. Based on observations made with the ESO Telescopes at the Paranal Observatory under programme ID 092.D-0420 and the NASA/ESA Hubble Space Telescope, obtained from the data archive at the Space Telescope Science Institute. STScI is operated by the Association of Universities for Research in Astronomy, Inc. under NASA contract NAS 5-26555. JDL and AJL acknowledge support from the UK Science and Technology Facilities Council (grant ID ST/I001719/1). AJL is grateful to the Leverhulme Trust for a Philip Leverhulme Prize award. RPC was supported by the Swedish Research Council (grant 2012-2254). MBD was supported by the Swedish Research Council (grants 2008-4089 and 2011-3991).

\newcommand{\araa}{ARA\&A}   \newcommand{\aap}{A\&A}
\newcommand{\aj}{AJ}         \newcommand{\apj}{ApJ}
\newcommand{\apjl}{ApJ}      \newcommand{\apjs}{ApJS}
\newcommand{\mnras}{MNRAS}   \newcommand{\nat}{Nature}
\newcommand{\pasj}{PASJ}     \newcommand{\pasp}{PASP}
\newcommand{\procspie}{Proc.\ SPIE} \newcommand{\physrep}{Phys. Rep.}
\newcommand{\apss}{APSS}
\newcommand{\solphys}{Sol. Phys.}
\newcommand{\actaa}{Acta Astronom}
\newcommand{\aaps}{A\&A Supp}
\newcommand{\iaucirc}{IAU Circular}
\bibliographystyle{mn2e}
\bibliography{/home/jdl/references}

\begin{thebibliography}{89}
\expandafter\ifx\csname natexlab\endcsname\relax\def\natexlab#1{#1}\fi

\bibitem[{{Anderson} {et~al}\mbox{.}(2012){Anderson}, {Habergham}, {James}, \&
  {Hamuy}}]{anderson12}
{Anderson} J.~P., {Habergham} S.~M., {James} P.~A., {Hamuy} M., 2012, \mnras,
  424, 1372

\bibitem[{{Anderson} \& {James}(2009)}]{anderson09}
{Anderson} J.~P., {James} P.~A., 2009, \mnras, 399, 559

\bibitem[{{Barbon} {et~al}\mbox{.}(1999){Barbon}, {Buond{\'{\i}}},
  {Cappellaro}, \& {Turatto}}]{barbon99}
{Barbon} R., {Buond{\'{\i}}} V., {Cappellaro} E., {Turatto} M., 1999, \aaps,
  139, 531

\bibitem[{{Berger}(2013)}]{berger13}
{Berger} E., 2013, arXiv:1311.2603

\bibitem[{{Bergeron}(2003)}]{bergeron03}
{Bergeron} P., 2003, \apj, 586, 201

\bibitem[{{Bertin} \& {Arnouts}(1996)}]{bertin96}
{Bertin} E., {Arnouts} S., 1996, \aaps, 117, 393

\bibitem[{{Bildsten} {et~al}\mbox{.}(2007){Bildsten}, {Shen}, {Weinberg}, \&
  {Nelemans}}]{bildsten07}
{Bildsten} L., {Shen} K.~J., {Weinberg} N.~N., {Nelemans} G., 2007, \apjl, 662,
  L95

\bibitem[{{Brown} {et~al}\mbox{.}(2010){Brown}, {Kilic}, {Allende Prieto}, \&
  {Kenyon}}]{brown10}
{Brown} W.~R., {Kilic} M., {Allende Prieto} C., {Kenyon} S.~J., 2010, \apj,
  723, 1072

\bibitem[{{Brown} {et~al}\mbox{.}(2011){Brown}, {Kilic}, {Allende Prieto}, \&
  {Kenyon}}]{brown11}
{Brown} W.~R., {Kilic} M., {Allende Prieto} C., {Kenyon} S.~J., 2011, \mnras,
  411, L31

\bibitem[{{Church} {et~al}\mbox{.}(2006){Church}, {Bush}, {Tout}, \&
  {Davies}}]{church06}
{Church} R.~P., {Bush} S.~J., {Tout} C.~A., {Davies} M.~B., 2006, \mnras, 372,
  715

\bibitem[{{Church} {et~al}\mbox{.}(2011){Church}, {Levan}, {Davies}, \&
  {Tanvir}}]{church11}
{Church} R.~P., {Levan} A.~J., {Davies} M.~B., {Tanvir} N., 2011, \mnras, 413,
  2004

\bibitem[{{Church} {et~al}\mbox{.}(in prep.)}]{churchprep}
{Church, R.~P.~et~al.}, in prep.

\bibitem[{{Courteau} {et~al}\mbox{.}(2011){Courteau}, {Widrow}, {McDonald},
  {Guhathakurta}, {Gilbert}, {Zhu}, {Beaton}, \& {Majewski}}]{courteau11}
{Courteau} S., {Widrow} L.~M., {McDonald} M., {Guhathakurta} P., {Gilbert}
  K.~M., {Zhu} Y., {Beaton} R.~L., {Majewski} S.~R., 2011, \apj, 739, 20

\bibitem[{{Crockett} {et~al}\mbox{.}(2007){Crockett}, {Smartt}, {Eldridge},
  {Mattila}, {Young}, {Pastorello}, {Maund}, {Benn}, \& {Skillen}}]{crockett07}
{Crockett} R.~M. {et~al.}, 2007, \mnras, 381, 835

\bibitem[{{Crowther}(2007)}]{crowther07}
{Crowther} P.~A., 2007, \araa, 45, 177

\bibitem[{{Davies}(1995)}]{davies95}
{Davies} M.~B., 1995, \mnras, 276, 887

\bibitem[{{Davies} \& {Benz}(1995)}]{davies95b}
{Davies} M.~B., {Benz} W., 1995, \mnras, 276, 876

\bibitem[{{Davies}, {Piotto} \& {de Angeli}(2004){Davies}, {Piotto}, \& {de
  Angeli}}]{davies04}
{Davies} M.~B., {Piotto} G., {de Angeli} F., 2004, \mnras, 349, 129

\bibitem[{{Davies}, {Ritter} \& {King}(2002){Davies}, {Ritter}, \&
  {King}}]{davies02}
{Davies} M.~B., {Ritter} H., {King} A., 2002, \mnras, 335, 369

\bibitem[{{Dessart} {et~al}\mbox{.}(2011){Dessart}, {Hillier}, {Livne}, {Yoon},
  {Woosley}, {Waldman}, \& {Langer}}]{dessart11}
{Dessart} L., {Hillier} D.~J., {Livne} E., {Yoon} S.-C., {Woosley} S.,
  {Waldman} R., {Langer} N., 2011, \mnras, 414, 2985

\bibitem[{{Drilling} \& {Landolt}(2000)}]{drilling00}
{Drilling} J.~S., {Landolt} A.~U., 2000, {Normal Stars}, New York: AIP Press;
  Springer, p. 381

\bibitem[{{Eldridge}, {Izzard} \& {Tout}(2008){Eldridge}, {Izzard}, \&
  {Tout}}]{eldridge08}
{Eldridge} J.~J., {Izzard} R.~G., {Tout} C.~A., 2008, \mnras, 384, 1109

\bibitem[{{Eldridge}, {Langer} \& {Tout}(2011){Eldridge}, {Langer}, \&
  {Tout}}]{eldridge11}
{Eldridge} J.~J., {Langer} N., {Tout} C.~A., 2011, \mnras, 414, 3501

\bibitem[{{Eldridge} \& {Stanway}(2009)}]{eldridge09}
{Eldridge} J.~J., {Stanway} E.~R., 2009, \mnras, 400, 1019

\bibitem[{{Filippenko} {et~al}\mbox{.}(2003){Filippenko}, {Chornock}, {Swift},
  {Modjaz}, {Simcoe}, \& {Rauch}}]{filippenko03}
{Filippenko} A.~V., {Chornock} R., {Swift} B., {Modjaz} M., {Simcoe} R.,
  {Rauch} M., 2003, \iaucirc, 8159, 2

\bibitem[{{Foley} {et~al}\mbox{.}(2013){Foley}, {Challis}, {Chornock},
  {Ganeshalingam}, {Li}, {Marion}, {Morrell}, {Pignata}, {Stritzinger},
  {Silverman}, {Wang}, {Anderson}, {Filippenko}, {Freedman}, {Hamuy}, {Jha},
  {Kirshner}, {McCully}, {Persson}, {Phillips}, {Reichart}, \&
  {Soderberg}}]{foley13}
{Foley} R.~J. {et~al.}, 2013, \apj, 767, 57

\bibitem[{{Fong} \& {Berger}(2013)}]{fong13b}
{Fong} W., {Berger} E., 2013, \apj, 776, 18

\bibitem[{{Fong} {et~al}\mbox{.}(2013){Fong}, {Berger}, {Chornock}, {Margutti},
  {Levan}, {Tanvir}, {Tunnicliffe}, {Czekala}, {Fox}, {Perley}, {Cenko},
  {Zauderer}, {Laskar}, {Persson}, {Monson}, {Kelson}, {Birk}, {Murphy},
  {Servillat}, \& {Anglada}}]{fong13}
{Fong} W. {et~al.}, 2013, \apj, 769, 56

\bibitem[{{Fong}, {Berger} \& {Fox}(2010){Fong}, {Berger}, \& {Fox}}]{fong10}
{Fong} W., {Berger} E., {Fox} D.~B., 2010, \apj, 708, 9

\bibitem[{{F{\"o}rster} \& {Schawinski}(2008)}]{forster08}
{F{\"o}rster} F., {Schawinski} K., 2008, \mnras, 388, L74

\bibitem[{{Fruchter} {et~al}\mbox{.}(2006){Fruchter}, {Levan}, {Strolger},
  {Vreeswijk}, {Thorsett}, {Bersier}, {Burud}, {Castro Cer{\'o}n},
  {Castro-Tirado}, {Conselice}, {Dahlen}, {Ferguson}, {Fynbo}, {Garnavich},
  {Gibbons}, {Gorosabel}, {Gull}, {Hjorth}, {Holland}, {Kouveliotou}, {Levay},
  {Livio}, {Metzger}, {Nugent}, {Petro}, {Pian}, {Rhoads}, {Riess}, {Sahu},
  {Smette}, {Tanvir}, {Wijers}, \& {Woosley}}]{fruchter06}
{Fruchter} A.~S. {et~al.}, 2006, \nat, 441, 463

\bibitem[{{Fryer} {et~al}\mbox{.}(2009){Fryer}, {Brown}, {Bufano}, {Dahl},
  {Fontes}, {Frey}, {Holland}, {Hungerford}, {Immler}, {Mazzali}, {Milne},
  {Scannapieco}, {Weinberg}, \& {Young}}]{fryer09}
{Fryer} C.~L. {et~al.}, 2009, \apj, 707, 193

\bibitem[{{Graham} {et~al}\mbox{.}(2003){Graham}, {Li}, {Puckett}, {Toth}, \&
  {Qiu}}]{graham03}
{Graham} J., {Li} W., {Puckett} T., {Toth} D., {Qiu} Y.~L., 2003, \iaucirc,
  8045, 1

\bibitem[{{Green}(2003)}]{IAUC8049}
{Green} D.~W.~E., 2003, \iaucirc, 8049, 3

\bibitem[{{Grindlay}, {Portegies Zwart} \& {McMillan}(2006){Grindlay},
  {Portegies Zwart}, \& {McMillan}}]{grindlay06}
{Grindlay} J., {Portegies Zwart} S., {McMillan} S., 2006, Nature Physics, 2,
  116

\bibitem[{{Gvaramadze}, {Gualandris} \& {Portegies Zwart}(2009){Gvaramadze},
  {Gualandris}, \& {Portegies Zwart}}]{gvaramadze09}
{Gvaramadze} V.~V., {Gualandris} A., {Portegies Zwart} S., 2009, \mnras, 396,
  570

\bibitem[{{Harris}(1991)}]{harris91}
{Harris} W.~E., 1991, \araa, 29, 543

\bibitem[{{Harris}(1996)}]{harris96}
{Harris} W.~E., 1996, \aj, 112, 1487

\bibitem[{{Ivanova} {et~al}\mbox{.}(2006){Ivanova}, {Heinke}, {Rasio}, {Taam},
  {Belczynski}, \& {Fregeau}}]{ivanova06}
{Ivanova} N., {Heinke} C.~O., {Rasio} F.~A., {Taam} R.~E., {Belczynski} K.,
  {Fregeau} J., 2006, \mnras, 372, 1043

\bibitem[{{Jarosik} {et~al}\mbox{.}(2011){Jarosik}, {Bennett}, {Dunkley},
  {Gold}, {Greason}, {Halpern}, {Hill}, {Hinshaw}, {Kogut}, {Komatsu},
  {Larson}, {Limon}, {Meyer}, {Nolta}, {Odegard}, {Page}, {Smith}, {Spergel},
  {Tucker}, {Weiland}, {Wollack}, \& {Wright}}]{jarosik11}
{Jarosik} N. {et~al.}, 2011, \apjs, 192, 14

\bibitem[{{Kasliwal} {et~al}\mbox{.}(2012){Kasliwal}, {Kulkarni}, {Gal-Yam},
  {Nugent}, {Sullivan}, {Bildsten}, {Yaron}, {Perets}, {Arcavi}, {Ben-Ami},
  {Bhalerao}, {Bloom}, {Cenko}, {Filippenko}, {Frail}, {Ganeshalingam},
  {Horesh}, {Howell}, {Law}, {Leonard}, {Li}, {Ofek}, {Polishook}, {Poznanski},
  {Quimby}, {Silverman}, {Sternberg}, \& {Xu}}]{kasliwal12}
{Kasliwal} M.~M. {et~al.}, 2012, \apj, 755, 161

\bibitem[{{Kasliwal} {et~al}\mbox{.}(2010){Kasliwal}, {Kulkarni}, {Gal-Yam},
  {Yaron}, {Quimby}, {Ofek}, {Nugent}, {Poznanski}, {Jacobsen}, {Sternberg},
  {Arcavi}, {Howell}, {Sullivan}, {Rich}, {Burke}, {Brimacombe},
  {Milisavljevic}, {Fesen}, {Bildsten}, {Shen}, {Cenko}, {Bloom}, {Hsiao},
  {Law}, {Gehrels}, {Immler}, {Dekany}, {Rahmer}, {Hale}, {Smith}, {Zolkower},
  {Velur}, {Walters}, {Henning}, {Bui}, \& {McKenna}}]{kasliwal10}
{Kasliwal} M.~M. {et~al.}, 2010, \apjl, 723, L98

\bibitem[{{Kawabata} {et~al}\mbox{.}(2010){Kawabata}, {Maeda}, {Nomoto},
  {Taubenberger}, {Tanaka}, {Deng}, {Pian}, {Hattori}, \&
  {Itagaki}}]{kawabata10}
{Kawabata} K.~S. {et~al.}, 2010, \nat, 465, 326

\bibitem[{{Kawka}, {Vennes} \& {Thorstensen}(2004){Kawka}, {Vennes}, \&
  {Thorstensen}}]{kawka04}
{Kawka} A., {Vennes} S., {Thorstensen} J.~R., 2004, \aj, 127, 1702

\bibitem[{{Kelly}, {Kirshner} \& {Pahre}(2008){Kelly}, {Kirshner}, \&
  {Pahre}}]{kelly08}
{Kelly} P.~L., {Kirshner} R.~P., {Pahre} M., 2008, \apj, 687, 1201

\bibitem[{{Kim} {et~al}\mbox{.}(2011){Kim}, {Kim}, {Hwang}, {Lee}, {Chun}, \&
  {Ann}}]{kim11}
{Kim} E., {Kim} M., {Hwang} N., {Lee} M.~G., {Chun} M.-Y., {Ann} H.~B., 2011,
  \mnras, 412, 1881

\bibitem[{{King}, {Olsson} \& {Davies}(2007){King}, {Olsson}, \&
  {Davies}}]{king07}
{King} A., {Olsson} E., {Davies} M.~B., 2007, \mnras, 374, L34

\bibitem[{{Kobulnicky} \& {Fryer}(2007)}]{kobulnicky07}
{Kobulnicky} H.~A., {Fryer} C.~L., 2007, \apj, 670, 747

\bibitem[{{Lada} \& {Lada}(2003)}]{lada03}
{Lada} C.~J., {Lada} E.~A., 2003, \araa, 41, 57

\bibitem[{{Li} {et~al}\mbox{.}(2003){Li}, {Filippenko}, {Chornock}, {Berger},
  {Berlind}, {Calkins}, {Challis}, {Fassnacht}, {Jha}, {Kirshner}, {Matheson},
  {Sargent}, {Simcoe}, {Smith}, \& {Squires}}]{li03}
{Li} W. {et~al.}, 2003, \pasp, 115, 453

\bibitem[{{Li} {et~al}\mbox{.}(2011){Li}, {Leaman}, {Chornock}, {Filippenko},
  {Poznanski}, {Ganeshalingam}, {Wang}, {Modjaz}, {Jha}, {Foley}, \&
  {Smith}}]{li11c}
{Li} W. {et~al.}, 2011, \mnras, 412, 1441

\bibitem[{{Lyman} {et~al}\mbox{.}(2014){Lyman}, {Bersier}, {James}, {Mazzali},
  {Eldridge}, {Fraser}, \& {Pian}}]{lyman14c}
{Lyman} J., {Bersier} D., {James} P., {Mazzali} P., {Eldridge} J., {Fraser} M.,
  {Pian} E., 2014, arXiv:1406.3667

\bibitem[{{Lyman}(2014)}]{lyman14b}
{Lyman} J.~D., 2014, PhD thesis, Liverpool John Moores University

\bibitem[{{Lyman} {et~al}\mbox{.}(2013){Lyman}, {James}, {Perets}, {Anderson},
  {Gal-Yam}, {Mazzali}, \& {Percival}}]{lyman13}
{Lyman} J.~D., {James} P.~A., {Perets} H.~B., {Anderson} J.~P., {Gal-Yam} A.,
  {Mazzali} P., {Percival} S.~M., 2013, \mnras, 434, 527

\bibitem[{{Mackey} {et~al}\mbox{.}(2006){Mackey}, {Huxor}, {Ferguson},
  {Tanvir}, {Irwin}, {Ibata}, {Bridges}, {Johnson}, \& {Lewis}}]{mackey06}
{Mackey} A.~D. {et~al.}, 2006, \apjl, 653, L105

\bibitem[{{Maoz}, {Mannucci} \& {Nelemans}(2013){Maoz}, {Mannucci}, \&
  {Nelemans}}]{maoz13}
{Maoz} D., {Mannucci} F., {Nelemans} G., 2013, arXiv:1312.0628

\bibitem[{{Massey}(2002)}]{massey02}
{Massey} P., 2002, \apjs, 141, 81

\bibitem[{{Mateo}(1998)}]{mateo98}
{Mateo} M.~L., 1998, \araa, 36, 435

\bibitem[{{Maund} \& {Smartt}(2005)}]{maund05}
{Maund} J.~R., {Smartt} S.~J., 2005, \mnras, 360, 288

\bibitem[{{McConnachie}(2012)}]{mcconnachie12}
{McConnachie} A.~W., 2012, \aj, 144, 4

\bibitem[{{Metzger}(2012)}]{metzger12}
{Metzger} B.~D., 2012, \mnras, 419, 827

\bibitem[{{Metzger} \& {Berger}(2012)}]{metzger12b}
{Metzger} B.~D., {Berger} E., 2012, \apj, 746, 48

\bibitem[{{Moriya} {et~al}\mbox{.}(2010){Moriya}, {Tominaga}, {Tanaka},
  {Nomoto}, {Sauer}, {Mazzali}, {Maeda}, \& {Suzuki}}]{moriya10}
{Moriya} T., {Tominaga} N., {Tanaka} M., {Nomoto} K., {Sauer} D.~N., {Mazzali}
  P.~A., {Maeda} K., {Suzuki} T., 2010, \apj, 719, 1445

\bibitem[{{Mulchaey}, {Kasliwal} \& {Kollmeier}(2014){Mulchaey}, {Kasliwal}, \&
  {Kollmeier}}]{mulchaey14}
{Mulchaey} J.~S., {Kasliwal} M.~M., {Kollmeier} J.~A., 2014, \apjl, 780, L34

\bibitem[{{Perets}(2014)}]{perets14}
{Perets} H.~B., 2014, arXiv:1407.2254

\bibitem[{{Perets} {et~al}\mbox{.}(2011){Perets}, {Gal-yam}, {Crockett},
  {Anderson}, {James}, {Sullivan}, {Neill}, \& {Leonard}}]{perets11}
{Perets} H.~B., {Gal-yam} A., {Crockett} R.~M., {Anderson} J.~P., {James}
  P.~A., {Sullivan} M., {Neill} J.~D., {Leonard} D.~C., 2011, \apjl, 728, L36

\bibitem[{{Perets} {et~al}\mbox{.}(2010){Perets}, {Gal-Yam}, {Mazzali},
  {Arnett}, {Kagan}, {Filippenko}, {Li}, {Arcavi}, {Cenko}, {Fox}, {Leonard},
  {Moon}, {Sand}, {Soderberg}, {Anderson}, {James}, {Foley}, {Ganeshalingam},
  {Ofek}, {Bildsten}, {Nelemans}, {Shen}, {Weinberg}, {Metzger}, {Piro},
  {Quataert}, {Kiewe}, \& {Poznanski}}]{perets10}
{Perets} H.~B. {et~al.}, 2010, \nat, 465, 322

\bibitem[{{Podsiadlowski}, {Joss} \& {Hsu}(1992){Podsiadlowski}, {Joss}, \&
  {Hsu}}]{pod92}
{Podsiadlowski} P., {Joss} P.~C., {Hsu} J.~J.~L., 1992, \apj, 391, 246

\bibitem[{{Pooley} {et~al}\mbox{.}(2003){Pooley}, {Lewin}, {Anderson},
  {Baumgardt}, {Filippenko}, {Gaensler}, {Homer}, {Hut}, {Kaspi}, {Makino},
  {Margon}, {McMillan}, {Portegies Zwart}, {van der Klis}, \&
  {Verbunt}}]{pooley03}
{Pooley} D. {et~al.}, 2003, \apjl, 591, L131

\bibitem[{{Poznanski} {et~al}\mbox{.}(2010){Poznanski}, {Chornock}, {Nugent},
  {Bloom}, {Filippenko}, {Ganeshalingam}, {Leonard}, {Li}, \&
  {Thomas}}]{poznanski10}
{Poznanski} D. {et~al.}, 2010, Science, 327, 58

\bibitem[{{Prieto}, {Stanek} \& {Beacom}(2008){Prieto}, {Stanek}, \&
  {Beacom}}]{prieto08}
{Prieto} J.~L., {Stanek} K.~Z., {Beacom} J.~F., 2008, \apj, 673, 999

\bibitem[{{Ruiz-Lapuente}(2014)}]{ruiz14}
{Ruiz-Lapuente} P., 2014, arXiv:1403.4087

\bibitem[{{Sana} {et~al}\mbox{.}(2012){Sana}, {de Mink}, {de Koter}, {Langer},
  {Evans}, {Gieles}, {Gosset}, {Izzard}, {Le Bouquin}, \& {Schneider}}]{sana12}
{Sana} H. {et~al.}, 2012, Science, 337, 444

\bibitem[{{Schlafly} \& {Finkbeiner}(2011)}]{schlafly11}
{Schlafly} E.~F., {Finkbeiner} D.~P., 2011, \apj, 737, 103

\bibitem[{{Sim} {et~al}\mbox{.}(2012){Sim}, {Fink}, {Kromer}, {R{\"o}pke},
  {Ruiter}, \& {Hillebrandt}}]{sim12}
{Sim} S.~A., {Fink} M., {Kromer} M., {R{\"o}pke} F.~K., {Ruiter} A.~J.,
  {Hillebrandt} W., 2012, \mnras, 420, 3003

\bibitem[{{Sirianni} {et~al}\mbox{.}(2005){Sirianni}, {Jee}, {Ben{\'{\i}}tez},
  {Blakeslee}, {Martel}, {Meurer}, {Clampin}, {De Marchi}, {Ford}, {Gilliland},
  {Hartig}, {Illingworth}, {Mack}, \& {McCann}}]{sirianni05}
{Sirianni} M. {et~al.}, 2005, \pasp, 117, 1049

\bibitem[{{Smith} \& {Maeder}(1989)}]{smith89}
{Smith} L.~F., {Maeder} A., 1989, \aap, 211, 71

\bibitem[{{Smith}(2014)}]{smith14}
{Smith} N., 2014, arXiv:1402.1237

\bibitem[{{Smith} {et~al}\mbox{.}(2011){Smith}, {Li}, {Filippenko}, \&
  {Chornock}}]{smith11b}
{Smith} N., {Li} W., {Filippenko} A.~V., {Chornock} R., 2011, \mnras, 412, 1522

\bibitem[{{Strolger} {et~al}\mbox{.}(2002){Strolger}, {Smith}, {Suntzeff},
  {Phillips}, {Aldering}, {Nugent}, {Knop}, {Perlmutter}, {Schommer}, {Ho},
  {Hamuy}, {Krisciunas}, {Germany}, {Covarrubias}, {Candia}, {Athey}, {Blanc},
  {Bonacic}, {Bowers}, {Conley}, {Dahl{\'e}n}, {Freedman}, {Galaz}, {Gates},
  {Goldhaber}, {Goobar}, {Groom}, {Hook}, {Marzke}, {Mateo}, {McCarthy},
  {M{\'e}ndez}, {Muena}, {Persson}, {Quimby}, {Roth}, {Ruiz-Lapuente},
  {Seguel}, {Szentgyorgyi}, {von Braun}, {Wood-Vasey}, \& {York}}]{strolger02}
{Strolger} L.-G. {et~al.}, 2002, \aj, 124, 2905

\bibitem[{{Sullivan} {et~al}\mbox{.}(2011){Sullivan}, {Kasliwal}, {Nugent},
  {Howell}, {Thomas}, {Ofek}, {Arcavi}, {Blake}, {Cooke}, {Gal-Yam}, {Hook},
  {Mazzali}, {Podsiadlowski}, {Quimby}, {Bildsten}, {Bloom}, {Cenko},
  {Kulkarni}, {Law}, \& {Poznanski}}]{sullivan11}
{Sullivan} M. {et~al.}, 2011, \apj, 732, 118

\bibitem[{{Tamm}, {Tempel} \& {Tenjes}(2007){Tamm}, {Tempel}, \&
  {Tenjes}}]{tamm07}
{Tamm} A., {Tempel} E., {Tenjes} P., 2007, arXiv:0707.4375

\bibitem[{{Tanvir} {et~al}\mbox{.}(2013){Tanvir}, {Levan}, {Fruchter},
  {Hjorth}, {Hounsell}, {Wiersema}, \& {Tunnicliffe}}]{tanvir13}
{Tanvir} N.~R., {Levan} A.~J., {Fruchter} A.~S., {Hjorth} J., {Hounsell} R.~A.,
  {Wiersema} K., {Tunnicliffe} R.~L., 2013, \nat, 500, 547

\bibitem[{{Tunnicliffe} {et~al}\mbox{.}(2014){Tunnicliffe}, {Levan}, {Tanvir},
  {Rowlinson}, {Perley}, {Bloom}, {Cenko}, {O'Brien}, {Cobb}, {Wiersema},
  {Malesani}, {de Ugarte Postigo}, {Hjorth}, {Fynbo}, \&
  {Jakobsson}}]{tunnicliffe14}
{Tunnicliffe} R.~L. {et~al.}, 2014, \mnras, 437, 1495

\bibitem[{{Valenti} {et~al}\mbox{.}(2014){Valenti}, {Yuan}, {Taubenberger},
  {Maguire}, {Pastorello}, {Benetti}, {Smartt}, {Cappellaro}, {Howell},
  {Bildsten}, {Moore}, {Stritzinger}, {Anderson}, {Benitez-Herrera}, {Bufano},
  {Gonzalez-Gaitan}, {McCrum}, {Pignata}, {Fraser}, {Gal-Yam}, {Le Guillou},
  {Inserra}, {Reichart}, {Scalzo}, {Sullivan}, {Yaron}, \& {Young}}]{valenti14}
{Valenti} S. {et~al.}, 2014, \mnras, 437, 1519

\bibitem[{{van den Bergh}(1994)}]{vandenbergh94}
{van den Bergh} S., 1994, \aj, 108, 2145

\bibitem[{{Verbunt} \& {Hut}(1987)}]{verbunt87}
{Verbunt} F., {Hut} P., 1987, in IAU Symposium, Vol. 125, The Origin and
  Evolution of Neutron Stars, {Helfand} D.~J., {Huang} J.-H., eds., p. 187

\bibitem[{{Waldman} {et~al}\mbox{.}(2011){Waldman}, {Sauer}, {Livne}, {Perets},
  {Glasner}, {Mazzali}, {Truran}, \& {Gal-Yam}}]{waldman11}
{Waldman} R., {Sauer} D., {Livne} E., {Perets} H., {Glasner} A., {Mazzali} P.,
  {Truran} J.~W., {Gal-Yam} A., 2011, \apj, 738, 21

\bibitem[{{Yuan} {et~al}\mbox{.}(2013){Yuan}, {Kobayashi}, {Schmidt},
  {Podsiadlowski}, {Sim}, \& {Scalzo}}]{yuan13}
{Yuan} F., {Kobayashi} C., {Schmidt} B.~P., {Podsiadlowski} P., {Sim} S.~A.,
  {Scalzo} R.~A., 2013, \mnras, 432, 1680

\end{thebibliography}

\label{lastpage}
\end{document}